\documentclass[11pt,preprint]{aastex}

\shorttitle{The optical counterpart of IRAS\,00470+6429}
\shortauthors{A.~S. Miroshnichenko et al.}

\begin{document}

\title{Toward Understanding The B[e] Phenomenon:\\ III. Properties of the optical counterpart
of IRAS\,00470+6429.\altaffilmark{1}\\}

\author{A.~S. Miroshnichenko\altaffilmark{2}, E.~L. Chentsov\altaffilmark{3},
V.~G. Klochkova\altaffilmark{3}, S.~V. Zharikov\altaffilmark{4},
K.~N. Grankin\altaffilmark{5}, A.~V. Kusakin\altaffilmark{6,7},
T.~L. Gandet\altaffilmark{8}, G. Klingenberg\altaffilmark{9}, S.
Kildahl\altaffilmark{10}, R.~J. Rudy\altaffilmark{11}, D.~K.
Lynch\altaffilmark{11,20}, C.~C. Venturini\altaffilmark{11}, S.
Mazuk\altaffilmark{11}, R.~C. Puetter\altaffilmark{12}, R.~B.
Perry\altaffilmark{13}, A.~C. Carciofi\altaffilmark{14}, K.~S.
Bjorkman\altaffilmark{15}, R.~O. Gray\altaffilmark{16}, S.
Bernabei\altaffilmark{17}, V.~F. Polcaro\altaffilmark{18}, R.~F.
Viotti\altaffilmark{18}, L. Norci\altaffilmark{19}\vskip 1.0cm}

\altaffiltext{1}{Partially based on data obtained at the 6--m BTA
telescope of the Russian Academy of Sciences, 3--m IRTF, 3--m Shane
telescope of the Lick Observatory, 2.7--m Harlan J. Smith and 2.1--m
Otto Struve telescopes of the McDonald Observatory, 2.1--m telescope
of the San Pedro Martir Observatory, 1.5--m telescope of the Loiano
Observatory, and 0.8--m telescope of the Dark Sky Observatory}

\altaffiltext{2}{Department of Physics and Astronomy, University of
North Carolina at Greensboro, Greensboro, NC 27402, USA,
a$\_$mirosh$@$uncg.edu}

\altaffiltext{3}{Special Astrophysical Observatory of the Russian
Academy of Sciences, Nizhnyj Arkhyz, 369167, Russia}

\altaffiltext{4}{Instituto de Astronom\'ia, Universidad Nacional
Aut\'onoma de M\'exico, Apartado Postal 877, 22830, Ensenada, Baja
California, M\'exico}

\altaffiltext{5}{Crimean Astrophysical Observatory, Nauchny, Crimea
334413, Ukraine}

\altaffiltext{6}{Sternberg Astronomical Institute, Universitetskij
pr. 13, Moscow, Russia}

\altaffiltext{7}{Fesenkov Astrophysical Institute, Kamenskoe plato,
Almaty 050020, Kazakhstan}

\altaffiltext{8}{Lizard Hollow Observatory, P.O. Box 89175, Tucson,
AZ 85752--9175, USA}

\altaffiltext{9}{Bossmo Observatory, Mo i Rana, Norway}

\altaffiltext{10}{Haugveien 1G 3300, Hokksund, Norway}

\altaffiltext{11}{The Aerospace Corporation, M2/266, P.O. Box 92957,
Los Angeles, CA 90009, USA}

\altaffiltext{12}{University of California, San Diego, 9500 Gilman
Dr., La Jolla, CA 92093, USA}

\altaffiltext{13}{Science Support Office, M/S 160, NASA Langley
Research Center, Hampton, VA 23681, USA}

\altaffiltext{14}{Instituto de Astronomia, Geofisica e Ciencias
Atmosfericas, Universidade de Sao Paulo, Rua do Matao 1226, Cidade
Universitaria, S\~ao Paulo, SP 05508--900, Brazil}

\altaffiltext{15}{Ritter Observatory, Department of Physics and
Astronomy, University of Toledo, Toledo, OH 43606--3390, USA}

\altaffiltext{16}{Department of Physics and Astronomy, Appalachian
State University, Boone, NC 28608, USA}

\altaffiltext{17}{INAF--Osservatorio Astronomico di Bologna, Via
Ranzani 1, 40127 Bologna, Italy}

\altaffiltext{18}{Istituto di Astrofisica Spaziale e Fisica Cosmica,
INAF, Via del Fosso del Cavaliere 100, 00133, Roma, Italy}

\altaffiltext{19}{School of Physical Sciences and NCPST, Dublin City
University, Glasnevin, Dublin 9, Ireland}

\altaffiltext{20}{Visiting Astronomer at the Infrared Telescope
Facility, which is operated by the University of Hawaii under
Cooperative Agreement no. NNX08AE38A with the National Aeronautics
and Space Administration, Science Mission Directorate, Planetary
Astronomy Program}

\begin{abstract}

FS\,CMa type stars are a group of Galactic objects with the
B[e] phenomenon. They exhibit strong emission-line spectra
and infrared excesses, which are most likely due to recently formed
circumstellar dust. The group content and identification criteria
were described in the first two papers of the series. In this paper
we report our spectroscopic and photometric observations of the
optical counterpart of IRAS\,00470+6429 obtained in 2003--2008. The
optical spectrum is dominated by emission lines, most of which have
P\,Cyg type profiles. We detected significant brightness variations,
which may include a regular component, and variable spectral line
profiles in both shape and position. The presence of a weak Li {\sc
I} 6708 \AA\ line in the spectrum suggests that the object is most
likely a binary system with a B2--B3 spectral type primary companion
of a luminosity $\log$ L/L$\odot$ = 3.9$\pm$0.3 and a late-type
secondary companion. We estimate a distance toward the object to be
2.0$\pm$0.3 kpc from the Sun.
\end{abstract}


\section{Introduction}\label{intro}

The B[e] phenomenon refers to the simultaneous presence of forbidden
lines and a strong near-infrared (IR) excess in the spectra of hot
stars \citep[typically late-O to early-A type,][]{as76}. The group
of FS\,CMa type stars consists of $\sim$40 Galactic objects with the
B[e] phenomenon that exhibit very strong emission-line spectra and a
steep decrease of the IR flux at $\lambda \ge 10-30 \mu$m
\citep{m07}. Until recently, the nature and evolutionary state of
the majority of these objects were considered controversial. Their
luminosity range (2.5$\le \log$ L/L$_{\odot} \le$4.5) indicates that
the stars are not very massive, while the location on the
Hertzsprung-Russell diagram (typically within the main sequence)
suggests that they are not extremely evolved. The shape of the IR
spectral energy distribution (SED) in combination with a lack of a
visible nebulosity suggest that circumstellar (CS) dust is compactly
distributed around the source of radiation. It has probably been
formed recently or is being formed now.

Furthermore, the very strong emission-line spectra of FS\,CMa type
objects are not expected from single stars, because theory predicts
relatively weak stellar winds for the above mentioned luminosity
range. Their H$\alpha$ line equivalent widths are typically over an
order of magnitude higher than those of other
intermediate-luminosity hot stars \citep[see, e.g., Be stars, Herbig
Ae/Be stars;][]{mir08}. It has been shown that a noticeable fraction
of them ($\sim$30\%) are either recognized or suspected binary
systems \citep{m07,m07a}. Mass transfer between the stellar
companions can be an important mechanism, responsible for
accumulation of large amounts of CS gas and formation of CS dust in
and around the system.

The CS matter distribution is a key factor that affects the observed
parameters. Strong and fast stellar winds (e.g., in OB supergiants)
introduce spectral line emission, but hardly alter the stellar
continuum radiation due to a steep outward decrease of the CS gas
density. In slowly expanding CS envelopes (e.g., Keplerian disks of
Be stars or slow winds of B[e] supergiants) the line emission can be
even stronger, and the free-free and bound-free radiation can
significantly increase the object's brightness in the optical
region. It has been shown that CS disks in Be stars can contribute
up to $\sim$70\% to the stellar optical continuum
\citep{telting93,carciofi06}. This effect can be important in
FS\,CMa type objects, and neglecting it may lead to an
overestimation of the luminosity.

Our previous studies of some FS\,CMa type objects \citep[see,
e.g.,][]{m00,m02} did not take into account the entire variety of
the CS effects on their parameters, because we had very sparse data
at that time. Over the last few years we obtained multiple
spectroscopic and photometric observations for a large fraction of
the group members. Beginning with this paper, we will try to study
carefully their properties in order to determine both stellar and
circumstellar characteristics and ultimately understand the reasons
for the B[e] phenomenon in the FS\,CMa group.

In this paper we report the results of our low- and high-resolution
optical and near-IR spectroscopy and optical photometry of the
optical counterpart of IRAS\,00470+6429. Initial information about
its spectrum and SED was reported by \citet{p06} and \citet{m07a}.
The object has been recognized as an emission-line star since the
1950's \citep{gg54} and repeatedly detected in later optical surveys
for H$\alpha$ line emitters \citep[see, e.g.,][]{cmc83,kw99}. It was
suggested to be a proto-planetary nebula (PPN) candidate by
\citet{meix99} and \citet{kh05}, but neither of these studies
confirmed this hypothesis (see Section \ref{discuss} for more
detail). The object is located in an H {\sc II} region S2--182 and
seems to be its only illumination source \citep{rag07}. No radio
flux exceeding 5 mJy was detected from the nebula at 4.89 GHz
\citep{fich93}. Based on a radial velocity of the CO emission from
the region ($v_{\rm lsr} = -$27 km\,s$^{-1}$), \citet{fb84}
estimated a distance toward it as 2.0$\pm$0.6 kpc. \citet{rus03}
gives a systemic radial velocity of $v_{\rm lsr} = -$32 km\,s$^{-1}$
(from various data in the radio region) and a distance of
1.4$\pm$0.5 kpc. Both distances depend on the adopted Galaxy
rotation curve and need verification with other data.

The paper has the following structure. The observations are
described in Section \ref{observ}, analysis of the observed
properties is presented in Section \ref{analysis}, discussion of the
nature and evolutionary state in Section \ref{discuss}, and results
are summarized in Section \ref{concl}.

\section{Observations} \label{observ}

\subsection{Spectra}\label{spectra}

Spectroscopic observations of IRAS\,00470+6429 were obtained at
several telescopes in 2003--2008. Information about the telescopes
and spectrographs we used is presented in Table \ref{t1}.

The optical spectrum of IRAS\,00470+6429 is dominated by emission
lines, most of which have P\,Cyg type profiles (mostly hydrogen and
singly ionized metals). In particular, it contains many Fe {\sc II}
lines, which is typical of an object with the B[e] phenomenon. The
absorption components of the P\,Cyg profiles are hard to trace,
because in most lines they are broad and weak. Other factors that
contribute to the problem are moderate signal-to-noise ratios at the
continuum level in our data (typically 20--100) and complicated
variations of the continuum in \'echelle spectra. There are many
pure emission lines, such as [O {\sc I}] 6300 and 6364 \AA, which
are the only forbidden lines detected in our spectra, weak and broad
He {\sc I} 5876 and 6678 \AA\ lines, and hydrogen lines of the
Paschen series in the red part of the spectrum. The O {\sc I}
triplet at 7770--7775 \AA\ exhibits either an emission-absorption
profile or a pure emission one. The near-IR Ca {\sc II} triplet
shows strong emission profiles. The equivalent width of the emission
component of the H$\alpha$ line varies between 60 \AA\ and 100 \AA\
with no obvious periodicity.

Absorption features detected in the spectrum are mostly interstellar
(Na {\sc I} 5889 and 5895 \AA\ and diffuse interstellar bands,
DIBs). Weak absorption lines of Mg {\sc II} 4481 \AA\ and Si {\sc
II} 5056 \AA, typical of a hot star, were detected in some of our
spectra. They are relatively narrow (full width at half maximum is
$\sim$50 km\,s$^{-1}$), suggesting a low rotation rate of the star.
A weak absorption line of Li {\sc I} 6708 \AA\ is clearly seen in
all the spectra obtained at McDonald (see Figure \ref{f2}). It is
the only detected feature of a late-type star that can be
interpreted a sign of the presence of a secondary companion (see
Section \ref{binary}). This line was also detected in the spectra of
several other FS\,CMa type objects \citep[see][]{m07a}.

\begin{figure}
\figurenum{1} \resizebox{\hsize}{!}{\includegraphics{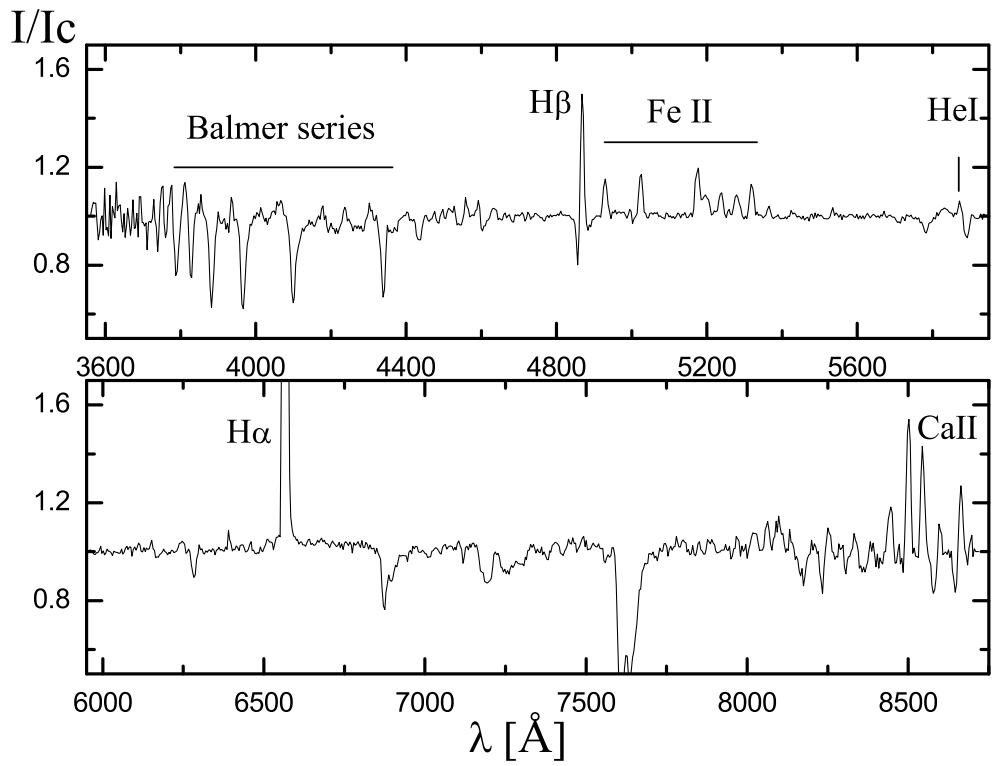}}
\caption{Low-resolution optical spectrum of IRAS\,00470+6429
obtained at Loiano Observatory. Intensities are normalized to the
nearby continuum and wavelengths are given in Angstr\"oms.
\label{f1}}
\end{figure}

\begin{figure}
\figurenum{2} \resizebox{\hsize}{!}{\includegraphics{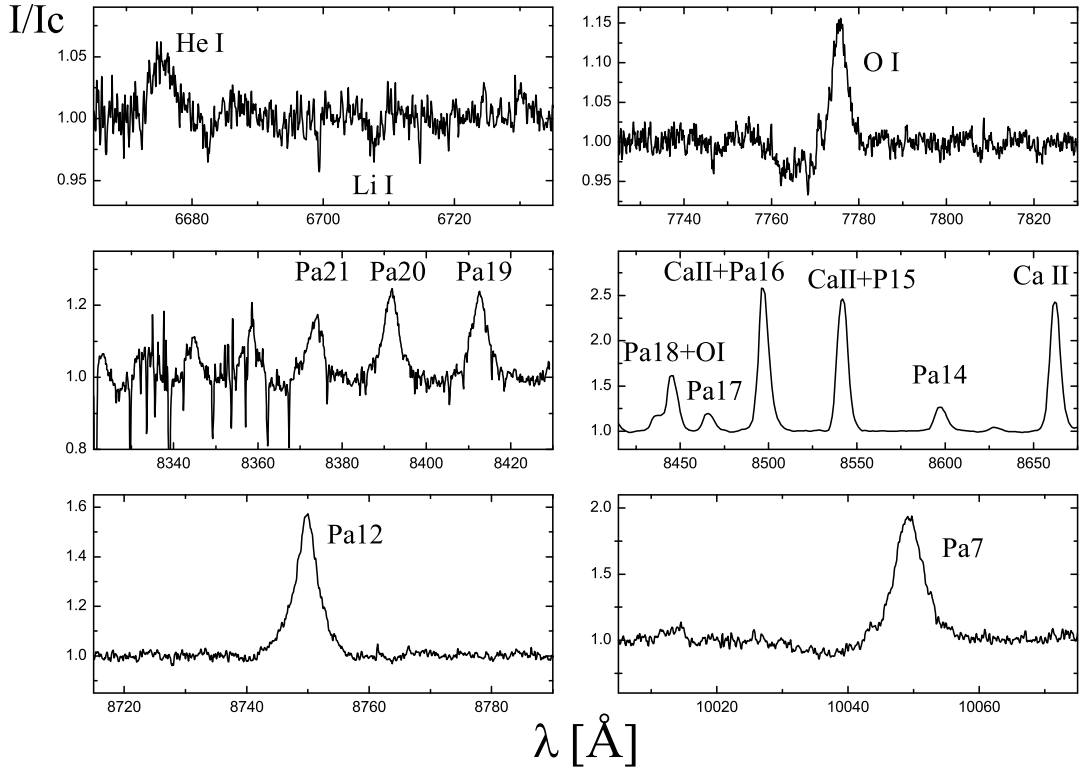}}
\caption{Most informative parts of the red portion of the spectrum
of IRAS\,00470+6429. The Ca {\sc II} triplet part (middle panel on
the right side) represents the 2007 December Lick spectrum. The
other five panels show the averaged high-resolution McDonald
spectrum obtained on 2006 December 26 and 27. Intensities and
wavelengths are in the same units as in Figure \ref{f1}. \label{f2}}
\end{figure}

\begin{figure}
\figurenum{3} \resizebox{\hsize}{!}{\includegraphics{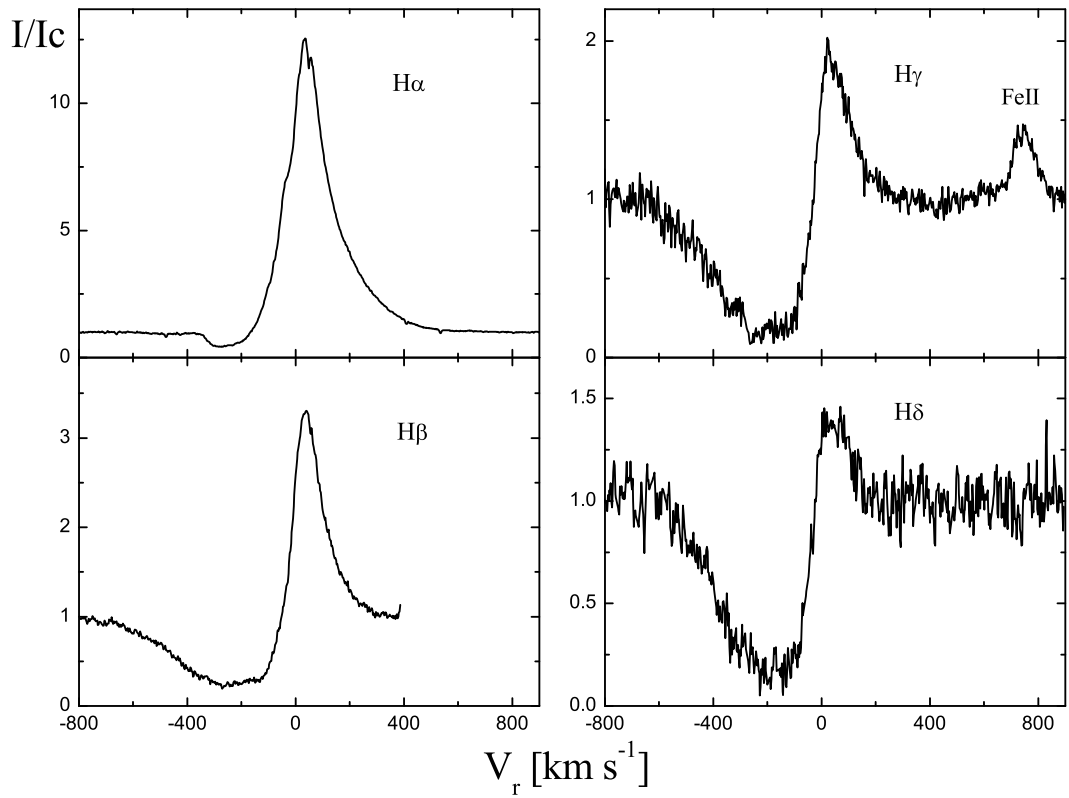}}
\caption{Balmer emission lines in the McDonald optical
high-resolution spectrum of IRAS\,00470+6429 obtained in December
2006. Intensities are normalized to the nearby continuum and
heliocentric radial velocities are given in km\,s$^{-1}$.
\label{f3}}
\end{figure}

The entire list of spectral lines detected in the object's optical
spectrum ($\lambda\lambda$ 0.38--1.05 $\mu$m) is presented in Table
\ref{t2}. The radial velocities of the most prominent species in the
spectra of some other stars close by on the sky that were observed
in order to help constrain the distance to the object, are listed in
Table \ref{t3}. Most features of the optical spectrum are shown in
Figure \ref{f1}-\ref{f3}.

\begin{figure}
\figurenum{4}
\begin{tabular}{cc}
\resizebox{9cm}{!}{\includegraphics{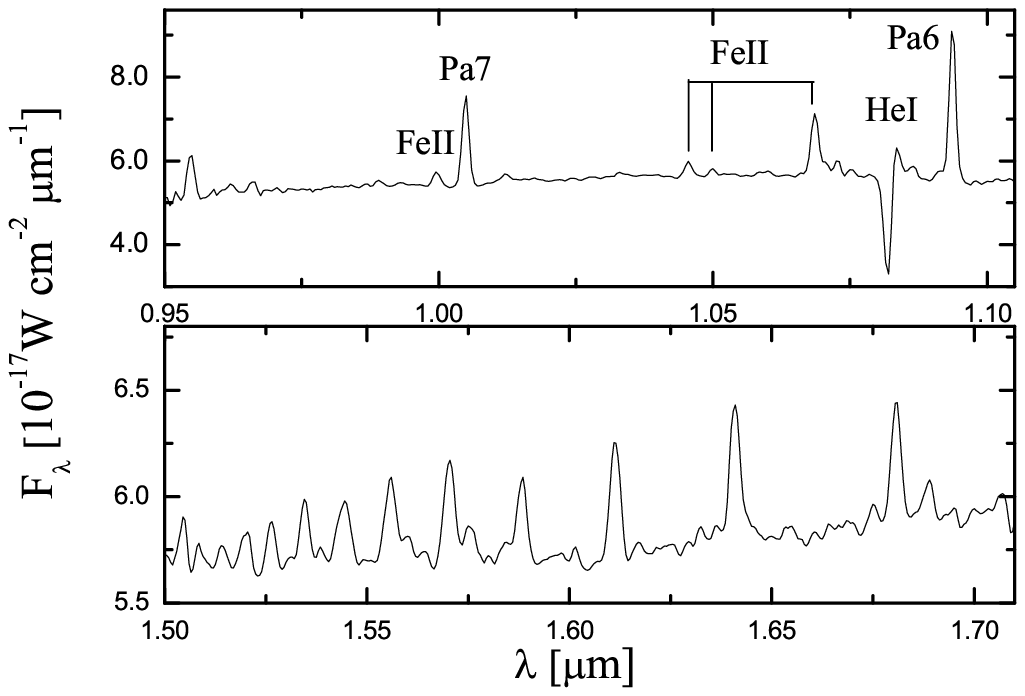}} &
\resizebox{8cm}{!}{\includegraphics{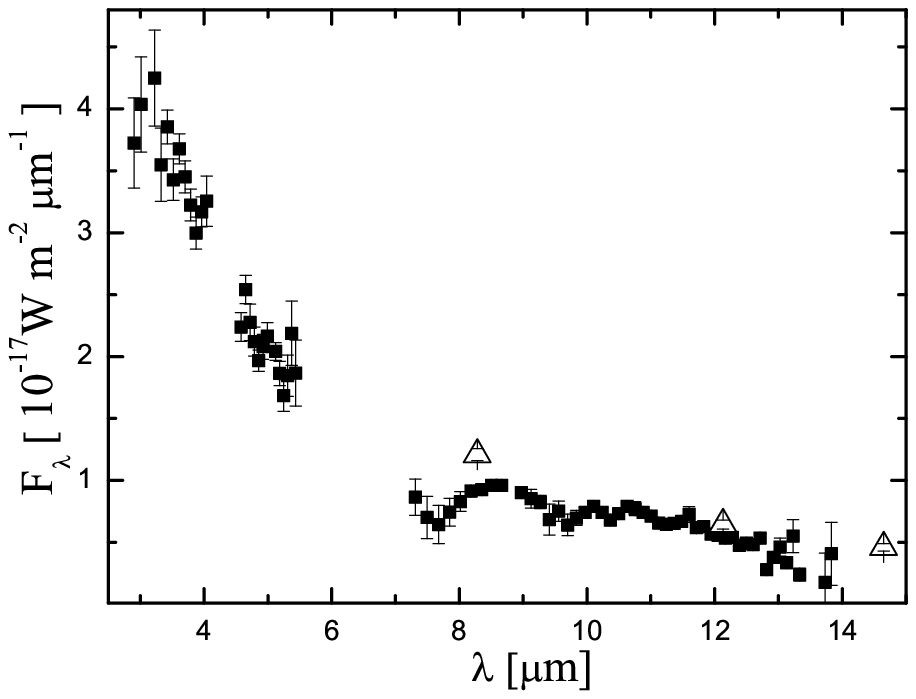}}\\
\end{tabular}
\caption{Left panel: Portions of the blue part of the near-IR
spectra of IRAS\,00470+6429 taken with the spectrometer NIRIS in
2003 December. Unmarked lines in the lower plot are those of the
Brackett series of hydrogen. Right panel: The BASS spectrum
(squares) and {\it MSX} data (triangles) of IRAS\,00470+6429.
Intensities are given in 10$^{-17}$ W\,m$^{-2}$\,$\mu$m$^{-1}$, and
wavelengths are given in microns. \label{f4}}
\end{figure}

The near-IR spectra were obtained at the 3-m Shane telescope of the
Lick Observatory with the Near-InfraRed Imaging Spectrograph
\citep[NIRIS,][]{rpm99} in 2003 December and 2007 December. The 2003
observations were taken in two channels, which provide nearly
continuous coverage between 0.8 and 2.5 $\mu$m. Additionally, an
optical channel was used in 2007 to cover a range between 0.48 and
0.9 $\mu$m. The average spectral resolution is $\sim$800 in the
optical channel and $\sim$700 in the near-IR channels. It shows only
emission lines (H {\sc I}, Fe {\sc II}, O {\sc I}) except for the He
{\sc I} 10830 \AA\ line, which has both emission and absorption
components (see the left panel of Figure \ref{f4}) and is different
from purely emission He {\sc I} lines in the optical region. The
Br$\gamma$ line with a maximum intensity of 1.4 relative to the
underlying continuum is weak \citep[shown by][]{m07a}, as it is
superimposed on the thermal continuum produced by CS dust.
\citet{kh05}, who took the object's spectrum in 1999 with a three
times lower resolving power, describe this line as strong at 1.2
continuum intensity at its maximum. The intensity difference can be
explained by the different spectral resolution.

Our near-IR spectra exhibit the same features with roughly the same
strengths, but the continuum level in 2007 is $\sim$20\% weaker than
that in 2003. The latter is consistent with the strong variations of
the optical continuum (see Figure \ref{f5}). The list of lines
detected in the near-IR region ($\lambda\lambda$ 1.05--2.5 $\mu$m)
is presented in Table \ref{t4}.

In 2003 October, IRAS\,00470+6429 was observed with the Broadband
Array Spectrograph \citep[BASS,][]{h90} at the 3--m IRTF in the
range 3--14 $\mu$m. The flux level in this spectral region is
consistent with that of the NIRIS data and the {\it MSX} data
\citep{egan03}. The BASS spectrum shows no noticeable features,
although there is some uncertainty in the region of the 9.7$\mu$m
silicate band due to difficulties in removing extinction by
atmospheric ozone. There is a change of slope in the continuum near
$\lambda \sim 7\mu$m (see right panel of Figure \ref{f4}), which is
probably due to the CS dust temperature structure.

\subsection{Photometry}\label{photometry}

The photometric observations were also obtained at several
observatories. At the Tien-Shan Observatory (Kazakhstan) and the
Maidanak Observatory (Uzbekistan) we used single-element
pulse-counting photometers. At the Lizard Hollow Observatory (LHO,
Tucson, AZ), Sonoita Research Observatory (AZ), and two
observatories in Norway (Bossmo and Persbuhaugen) we used CCD
photometers. All our long-term sets taken at were reduced to the
standard Johnson (Maidanak) or Johnson-Cousins (LHO and Sonoita)
photometric system. The log of the photometric observations is
presented in Table \ref{t5}.

Before our campaign, the object was observed in the course of the
Northern Star Variability Survey \citep[NSVS,][]{w04a} with no
filter. To compare these data with our $V$-band photometry, we used
a transformation derived by \citet[][$m_{v}(nsvs) = m_{v}(tycho) -
(m_{b}(tycho) - m_{v}(tycho))/1.875$]{w04a} and the average $B-V$
color from our Maidanak data. Since this transformation is based on
the Tycho photometric system, which slightly deviates from the
standard Johnson system \citep{hog98}, we expect a small systematic
shift between the corrected NSVS and our data. The object's light
curve in the $V$--band is shown in Figure \ref{f5}.

We have used the northern location of the object to monitor its
short-term variations during long winter nights at two observatories
in Norway. In total we obtained three series of observations over
seven hours each. The results are shown in Figure \ref{f6}. The
object's brightness shows no significant variations on a timescale
of a few hours. The $\sim$0.06--mag fading on 2007 January 23 and a
similar brightening on 2006 December 28 may be due to changes in the
CS medium, which is also suggested by the spectral line variations
(see Section \ref{spectra}).

In all the observations we used BD+63$^{\circ}$98 as a comparison
star (at 2$\arcmin$ from the object) and USNO--B1.0 1547--0023650
(at 4$\arcmin$) as a check star. No relative variations exceeding
0.02 mag have been detected for this pair of stars.

\section{Data analysis}\label{analysis}

\subsection{Light curve}\label{lc}

The light curve was analyzed using the PERIOD software time-series
analysis package from the Starlink Software
Collection\thanks{Information about the collection is available from
http://starlink.jach.hawaii.edu/release/}. Periods were searched
with two different methods: the $\chi^2$ method and the Lomb-Scargle
periodogram analysis \citep{scargle82}. The first method is a
straightforward technique, in which the input data are folded on a
series of trial periods. At each trial period, the data are fitted
with a sinusoid. The resulting reduced-$\chi^2$ values are plotted
as a function of the trial frequency, and the minima suggest the
most likely periods. This method is ideally suited to study any
sinusoidal variations \citep{hws86}. The second method is a novel
type of periodogram analysis, quite powerful for finding and testing
the significance of weak periodic signals in otherwise random,
unevenly sampled data \citep[see][]{hb86,pr89}.

False-alarm probabilities (FAP1) for the period were computed using
a Fisher randomization test or a Monte-Carlo method \citep[see,
e.g.,][]{nn85}. FAP1 represents the fraction of permutations (i.e.
shuffled time-series) that contained a trough lower than (in the
case of the $\chi^2$ method) or a peak higher than (in the case of
the Lomb-Scargle periodogram analysis) that of the periodogram of an
unrandomized dataset at any frequency. Therefore, this represents a
probability that, given the frequency search parameters, no periodic
component with this frequency is present in the data. To ensure
reliable significance values, the minimum number of permutations was
set to 1000. If a FAP1 lies between 0.00 and 0.01, then the
corresponding period is correct with 95\% confidence. Periods were
searched for within an interval ranging from 0.5 day to 1000 days.
The periodogram is computed at 1000 equally spaced frequencies
between 0.001 and 2. As these two methods have yielded identical
results, we discuss below only results of the Lomb-Scargle method.

\begin{figure}
\figurenum{5} \begin{tabular}{cc}
\resizebox{8.2cm}{!}{\includegraphics{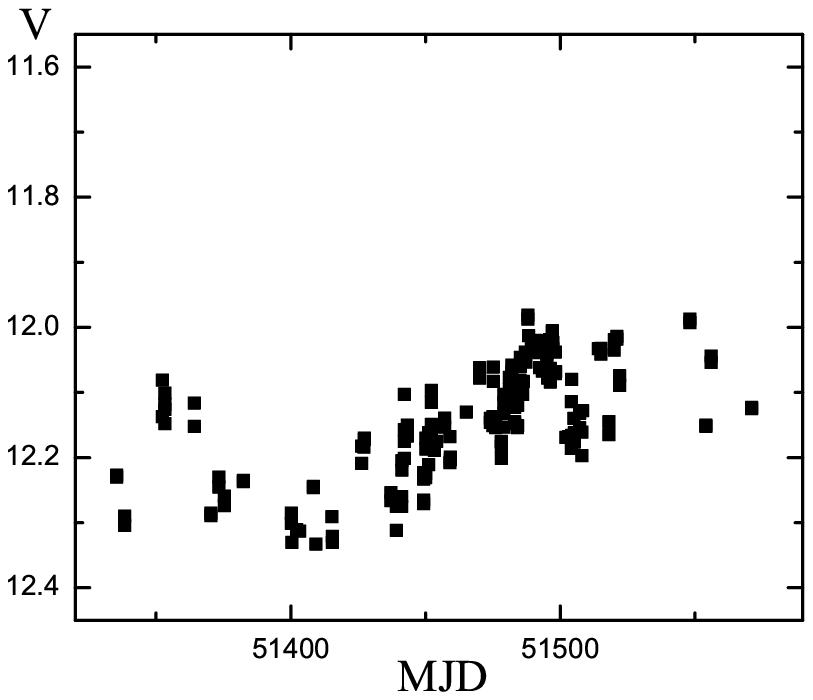}} &
\resizebox{8.8cm}{!}{\includegraphics{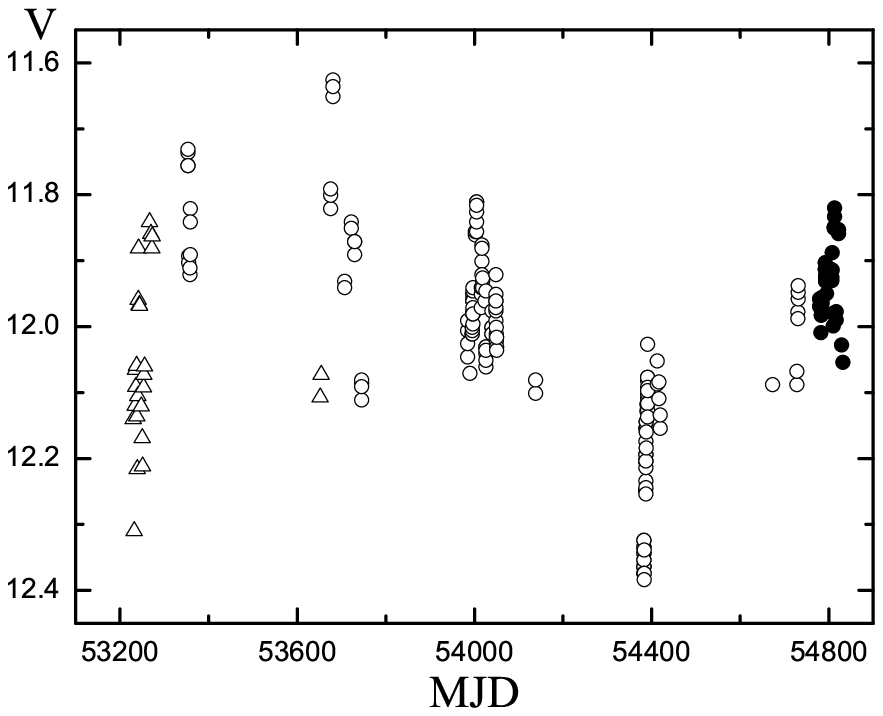}}\\
\end{tabular}
\caption{Optical light curve of IRAS\,00470+6429. The left panel
shows the NSVS data corrected for the color effect (see Section
\ref{photometry}). The right panel shows our observations: open
triangles, Maidanak data; open circles, LHO data; filled circles,
SRO data. Time is given in modified Julian dates (JD$-$2400000).
\label{f5}}
\end{figure}

\begin{figure}
\figurenum{6} \resizebox{14cm}{!}{\includegraphics{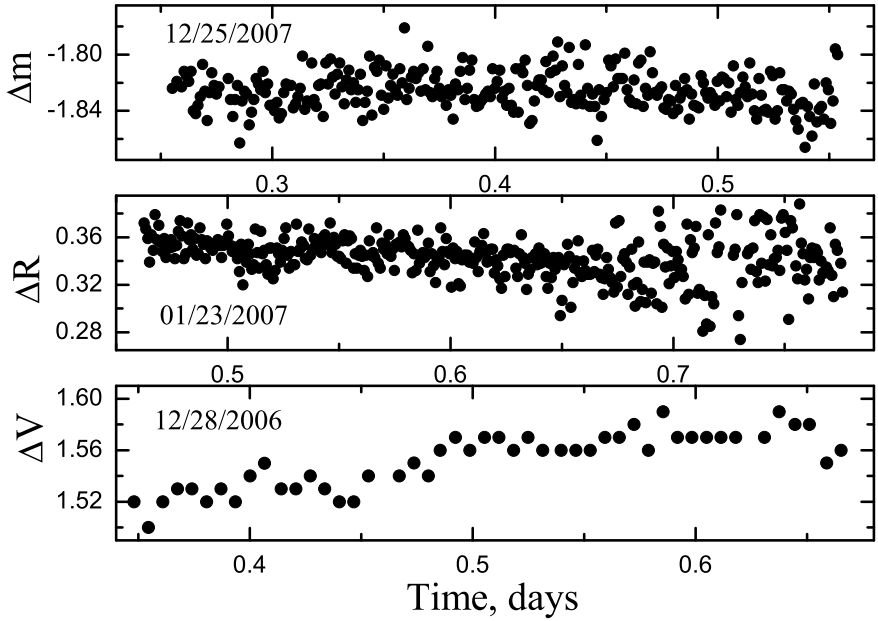}}
\caption{Short-term variations of the optical brightness of
IRAS\,00470+6429. The lower panel shows our Persbuhaugen data, while
the top two panels show our Bossmo data. For the data description
see Table \ref{t5}. \label{f6}}
\end{figure}

\begin{figure}
\figurenum{7} \resizebox{14cm}{!}{\includegraphics{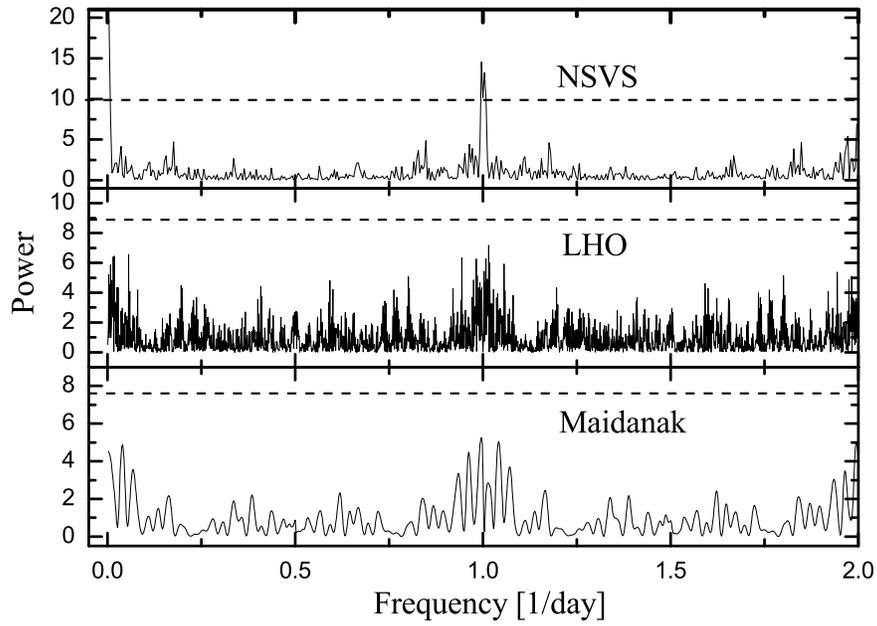}}
\caption{Power spectra of the periodograms for the data from
different observatories (see Table \ref{t5}). Dashed lines show a
99\% confidence level for peaks to be considered real. The level was
calculated using a prescription by \citet{hb86}.\label{f7}}
\end{figure}

\begin{figure}
\figurenum{8} \resizebox{12cm}{!}{\includegraphics{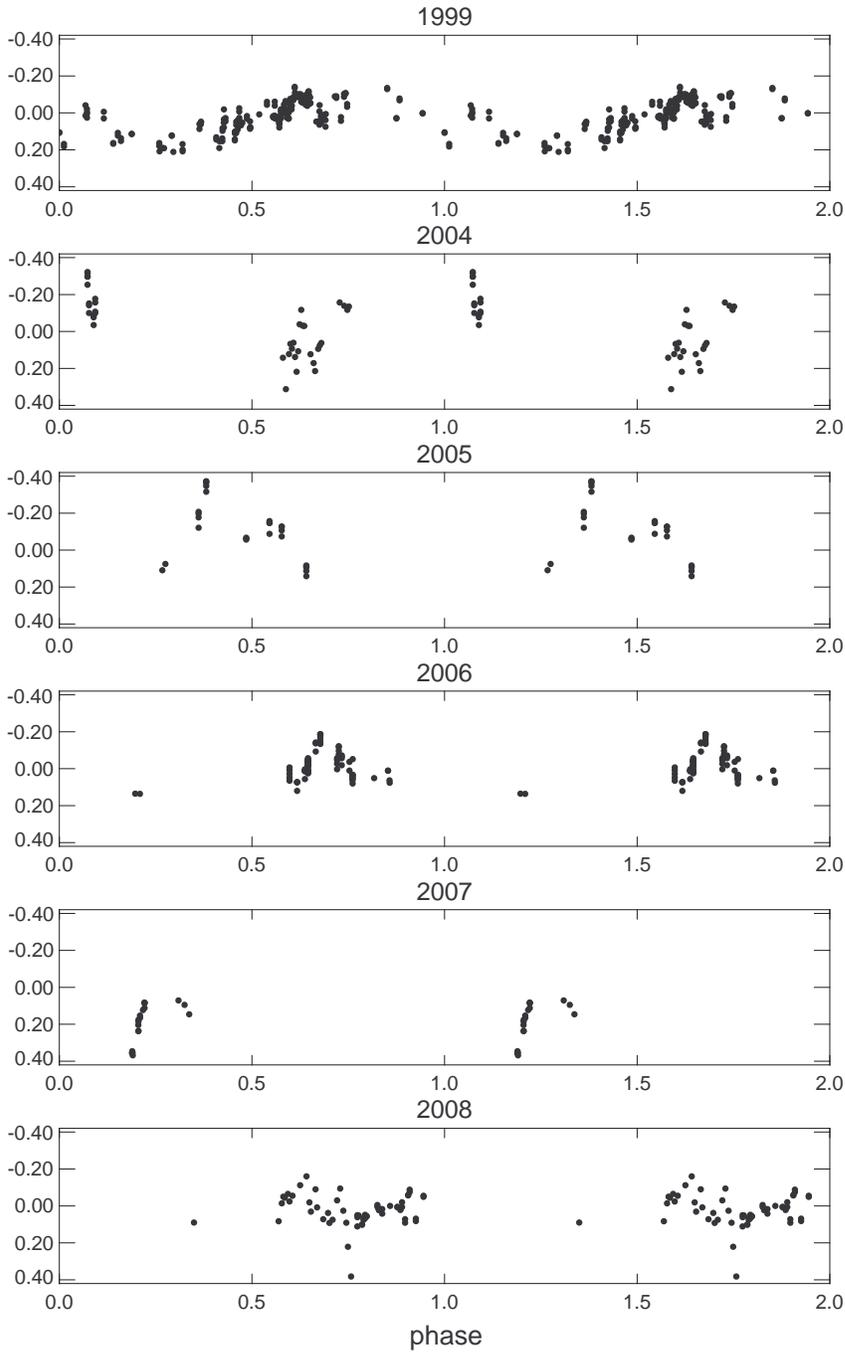}}
\caption{The $V$-band light curve of IRAS\,00470+6429 folded with
the 250--day period. The observed brightness is plotted with respect
to the average value, determined separately for the NSVS dataset
(the upper panel) and for our own observations (other panels). The
phases are calculated with respect to the first observing date of
the NSVS dataset.\label{f8}}
\end{figure}

We analyzed each of the long-term datasets separately as well as the
combined one. We found that the Maidanak and LHO datasets show no
significant cycles. The NSVS data exhibit two cycles, whose
durations are $\sim$1 day and 250$\pm$20 days (Figure \ref{f7}). A
careful analysis shows that the shorter cycle is caused by on-off
time ratio of observations close to one day. The 250--day cycle can
be easily recognized in the left panel of Figure \ref{f5}.
Unfortunately, the length of the NSVS dataset (only 237 days)
permits no precise estimate the cycle duration. For example,
\citet{w04b} report a 302--day cycle for the same NSVS dataset. An
analysis of own photometric data (LHO + Maidanak + SRO) obtained in
2004--2008 as well as of the combined dataset confirms neither of
these long-term cycles. We should note that this might be partially
due to a highly unequal time distributions of observations in our
datasets (especially in the LHO data). There are only 60 nights with
brightness measurements during a total period of 1427 nights. This
leads to a low significance level of the peaks in the power
spectrum. A $V$-band light curve of IRAS\,00470+6429 for the
combined dataset folded with the 250--day period is shown in Figure
\ref{f8}.

Another cycle that emerges from our data is based on two minima near
JD2453200 and JD2454400 (Figure \ref{f5}). This timescale is close
to that between the two minima of the He {\sc I} line radial
velocity, which occurred near JD2452900 and JD2454100 (Figure
\ref{f11}). The $\sim$1200--day cycle is probably of CS origin,
because we observe no strong radial velocity variations of the Li
{\sc I } line, which should trace the motion of a secondary
companion (see the discussion in Section \ref{binary}). Further
photometric and spectroscopic monitoring is needed to verify whether
the cycle is stable.

\subsection{Distance}\label{distance}

The position of \object{IRAS\,00470+6429} in the sky is projected
onto a region, occupied by the association \object{Cas\,OB7}. The
association is thought to be located in the Perseus spiral arm, but
the distance toward it is uncertain. It contains $\sim$40 B-type
stars, whose brightness is dimmed by interstellar extinction in a
range of A$_V$ from $\sim$2 to $\sim$3 mag. \citet{hump78} placed it
at 2.5 kpc from the Sun by averaging distances, calculated from the
luminosity and reddening of individual members. \citet{gs92} adopted
a distance of 1.8 kpc, based on the best fit to an intrinsic
color-magnitude diagram for the main-sequence B-type stars.
\citet{cp03} argued that Cas\,OB7 is surrounded by an expanding H
{\sc I} shell, whose approaching side is missing due to expansion
into interarm space. These authors studied interstellar gas and dust
around the association, and concluded that it is very patchy; and
adopted a distance of 2 kpc for the association.

\begin{figure}
\figurenum{9} \resizebox{12cm}{!}{\includegraphics{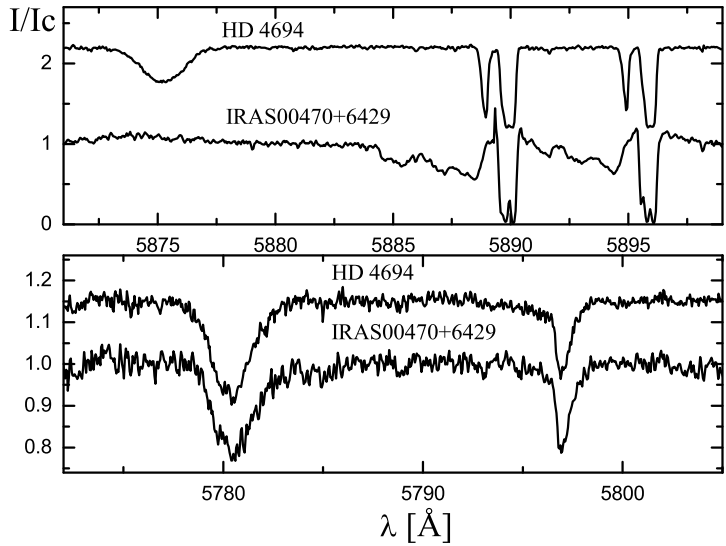}}
\caption{Portions of the optical high-resolution spectra of
IRAS\,00470+6429 and HD\,4694 that contain interstellar features.
Both spectra were obtained on the same night at the 2.7-m Harlan J.
Smith telescope in 2006 December. The upper panel shows the He {\sc
I} 5876 \AA\ line and the Na {\sc I} D-lines. The lower panel shows
the DIBs at 5780 \AA\ and 5797 \AA. Intensities and wavelengths are
in the same units as in Figure \ref{f1}. \label{f9}}
\end{figure}

In order to constrain the distance of IRAS\,00470+6429, we collected
available photometric and spectroscopic information about
projectionally close hot stars and obtained our own high-resolution
spectra of six members of Cas\,OB7 at angular distances of $\le
1\degr$ from the object (see Table \ref{t3}). Studying our spectra,
we found a striking difference in the structure of the interstellar
components of the Na {\sc I} D-lines for the stars with very similar
spectroscopic distances and reddenings. Some of them
(\object{Hiltner\,74} and BD+63$\degr$87) exhibit only one
absorption component in each of the D-lines, while the others
(Hiltner\,62, HD\,4694, HD\,4841, and BD+63$\degr$102) exhibit two
components. The low-velocity component at $-$(11--18) km\,s$^{-1}$
originates in the local arm, while the high-velocity one at
$-$(58--66) km\,s$^{-1}$ most likely forms in the Perseus arm within
the association. The absence of the high-velocity component in the
spectrum of IRAS\,00470+6429 in combination with a high reddening
and a strength of the diffuse interstellar bands, which is virtually
equal to that of HD\,4694 (see the lower panel of Figure \ref{f9}),
BD+63$\degr$102, and BD+63$\degr$87, suggests that our object is
located near the closer edge of both the association and the Perseus
arm. The high-velocity interstellar component of the sodium lines in
the spectrum of IRAS\,00470+6429, if it exists, is weak or blends
with the CS absorption components (see the upper panel of Figure
\ref{f9}).

The position and width of the Na {\sc I} D-line profiles in the
spectrum of IRAS\,00470+6429 are also very close to those of
\object{$\kappa$ Cas}, a supergiant from the association
\object{Cas\,OB14} \citep{hump78}. Cas\,OB14 is located at $\sim
3\degr$ from Cas\,OB7, but Cas\,OB14 is about half as far from the
Sun, and its stars are much less reddened than those of Cas\,OB7
(see Figure \ref{f10}). Additionally, heliocentric radial velocities
of the emission components of the metallic lines in the spectrum of
IRAS\,00470+6429 ($-$(40--20) km\,s$^{-1}$) are closer to those of
members of Cas\,OB7 ($-$(60--35) km\,s$^{-1}$) than to those of
members of Cas\,OB14 \citep[e.g., $-$(2--9) km\,s$^{-1}$ for
HD\,2905,][]{m57,hump78}.

The spectroscopic distances of the Cas\,OB7 members with one Na {\sc
I} D-line component are $\sim$0.1 kpc smaller than those of the
members with two line components (see Table \ref{t3}). Taking into
account uncertainties in the distance determination due to using
average luminosities for the adopted MK types \citep{sk81}, we
estimate a distance of 2.0$\pm$0.3 kpc toward IRAS\,00470+6429, as
an initial approximation. Its lower limit is constrained by the
distance of the nearest edge of the Perseus arm, which is not
precisely known. Its upper limit cannot exceed distances toward
stars with two sodium line components near the object. The distance
toward IRAS\,00470+6429 is also in agreement with independently
derived distance of Sh2--182 (see Section \ref{intro}).

High-resolution spectroscopy of more members of Cas\,OB7 is needed
to put better constraints on the distances. We also note that the
distance of Cas\,OB7 adopted by \citet{gs92} does not appear
well-justified, as not all stars in it might have been born at the
same time. This opinion is corroborated by large deviations of
spectroscopic distances of individual stars from 1.8 kpc.

Our observations of stars near IRAS\,00470+6429 also revealed that
radial velocities of Hiltner\,62 and BD\,+63$\arcdeg$87 are
inconsistent with the Galactic rotation for their spectroscopic
distances (see Table \ref{t3}), probably indicating their binary
nature. Based on our spectra, we classify HD\,4694 as B3 {\sc I}b
and BD+63$\degr$102 as B1 {\sc III} instead of B3 {\sc I}a and B1
{\sc II}, respectively, as suggested earlier \citep[see,
e.g.,][]{gs92}.

\subsection{Stellar parameters} \label{stel_par}

The average brightness of IRAS\,00470+6429 ($V = 12.0\pm0.3$ mag)
and the above distance estimate rule out a high luminosity for the
hot star. The absence of luminosity sensitive photospheric lines,
such as Si {\sc III} 5739 \AA\ \citep{m04} and Si {\sc II} 6347 and
6371 \AA\ \citep{ros74}, which are seen in the spectra of
supergiants with even stronger line emission, supports this
conclusion.

\begin{figure}
\figurenum{10} \resizebox{10cm}{!}{\includegraphics{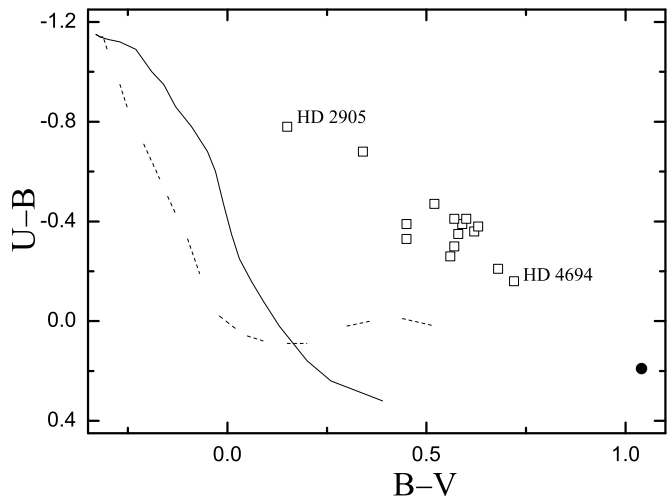}}
\caption{Color-color diagram for the stars in the region of the sky
within $\sim 1\degr$ from IRAS\,00470+6429 (HD\,2905, which is at
$\sim 3\degr$ from it, is shown for comparison of the reddening).
The object is shown by the filled circle. The solid line represents
the intrinsic color-indices of supergiants, while the dashed line
shows color-indices of dwarfs. \label{f10}}
\end{figure}

Our multicolor photometric data give the following average
color-indices in the Johnson photometric system: $\overline{U-B} =
0.19\pm0.04$, $\overline{B-V} = 1.04\pm0.03$, and $\overline{V-R} =
1.16\pm0.03$ mag. Comparison with those of projectionally close
B-type stars from Cas\,OB7 and Cas\,OB14 shows that IRAS\,00470+6429
has the strongest reddening. Part of it is definitely due to the CS
matter, because even more distant stars of Cas\,OB7 are bluer than
the object (see Figure \ref{f10}). Using a relationship between the
DIB strengths and reddening from \citet{h93}, we estimate the
interstellar component of the reddening toward IRAS\,00470+6429 to
be $E(B-V) \sim 0.85$ mag. This estimate coincides with that
calculated for HD\,4694 \citep[$B-V$ = 0.72 mag,][]{hiltner56}, thus
supporting the use of the relationship.

The CS component of the reddening depends somewhat on the object's
spectral type, which can be constrained from the spectral line
content. The absence of high excitation lines, the weakness of the
He {\sc I} emission lines (which may be excited by a different
source, see Section \ref{gas}), and the average optical
color-indices (assuming that the wavelength dependence of the CS
reddening is not very different from that of the interstellar one)
suggest that the star's spectral type is B2--B3. Therefore, the CS
part of the reddening is $E(B-V) \sim 0.35$ mag.

Assuming that the overall selective reddening is $E(B-V) \sim 1.2$
mag, the total extinction is A$_V$ = 3.1$\times E(B-V) \sim$ 3.8 mag,
the star's effective temperature is T$_{\rm eff} \sim$ 20000 K, and
the distance is 2.0$\pm$0.3 kpc, one can estimate the star's
luminosity to be $\log$ (L/L$\odot$) = 3.9$\pm$0.3. This estimate
places IRAS\,00470+6429 in the middle of the luminosity distribution
of the FS\,CMa group \citep{m07}.

The adopted luminosity is somewhat uncertain, because neither a
contribution to the optical continuum from the CS gas nor a possible
deviation of the CS reddening law from the interstellar one was
taken into account. This can only be resolved by careful modeling,
which will be presented in the next paper of this series. However,
we do not expect the correction to be large, because we still detect
photospheric lines (see Section \ref{spectra}). They are weak,
indicating that the photospheric spectrum is partially veiled by the
CS free-free and bound-free continuum radiation. Nevertheless, the
observed optical color-indices of IRAS\,00470+6429 do not noticeably
deviate from those expected from a normal B2--B3 star affected by
interstellar reddening (see Figure \ref{f10}).

\subsection{CS gas}\label{gas}

The presence of P\,Cyg profiles in the spectrum of IRAS\,00470+6429
is unusual for the FS\,CMa star group, the majority of whose members
exhibit single- or double-peaked profiles that are typical for
disk-like CS gaseous envelopes. There are only two other objects in
the entire group with P\,Cyg type profiles: \object{AS\,78}
\citep{m00} and \object{HD\,85567} \citep{m01}. However, this type
of a line profile does not exclude a disk-like geometry, which is
natural in a binary system \citep[our working hypothesis for
explaining properties of FS\,CMa stars,][]{m07}. In this case, the
orbital plane should be viewed nearly edge-on. In combination with
narrow photospheric lines, this suggestion favors a low rotational
velocity of the primary companion.

P\,Cyg type profiles are observed in various groups of early-type
emission-line stars. In supergiants with accelerating stellar winds,
spectral lines are not very strong because of a fast matter density
decrease with distance from the star. The position of the blue edge
of the line absorption component roughly marks the wind terminal
velocity. In pre-main-sequence Herbig Ae/Be stars with more
complicated CS envelopes, which include both accretion (in a disk)
and outflow (in a wider range of angles), the material is thought to have
higher velocities near the star \citep[for a review and an example
see][]{bp04}. Therefore, the blue edge of the line may mark the
velocity of material that was recently ejected from the stellar
surface. It seems unlikely that the wind of a relatively
low-luminosity star, such as IRAS\,00470+6429, will accelerate up to
the radial velocities of the blue edges of the Balmer lines
(in some spectra they reach $\sim$900 km\,s$^{-1}$) through
radiation pressure in spectral lines as in supergiants.

Thus, we assume a different mechanism of the CS gaseous envelope
formation in the object. Following the reasoning of \citet{m07},
IRAS\,00470+6429 does not seem to be a young star, and one should
not expect a protostellar type accretion to be responsible for the
formation of the CS disk. On the other hand, the object is probably
a binary system (see Section \ref{binary}), in which the disk forms
through mass loss from the hot primary or mass transfer from the
secondary. Additional matter ejection in the line of sight, even if
the orbital plane deviates from this direction, could account for
the observed absorption components of the P\,Cyg line profiles.

\begin{figure*}
\figurenum{11}
\begin{tabular}{cc}
\resizebox{8.8cm}{!}{\includegraphics{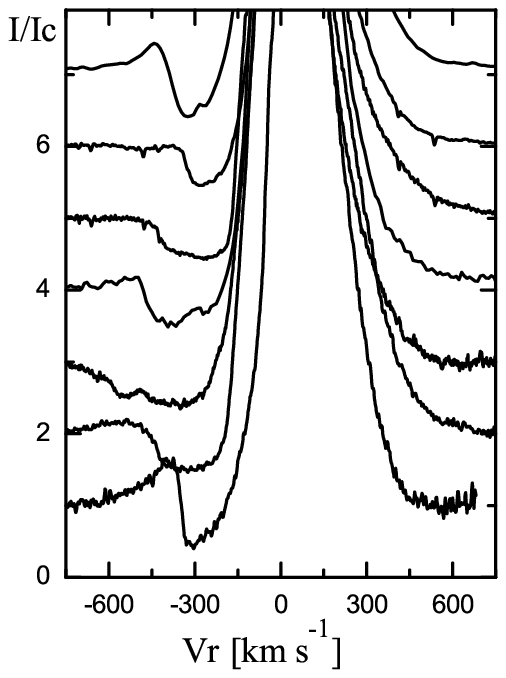}} &
\resizebox{8.5cm}{!}{\includegraphics{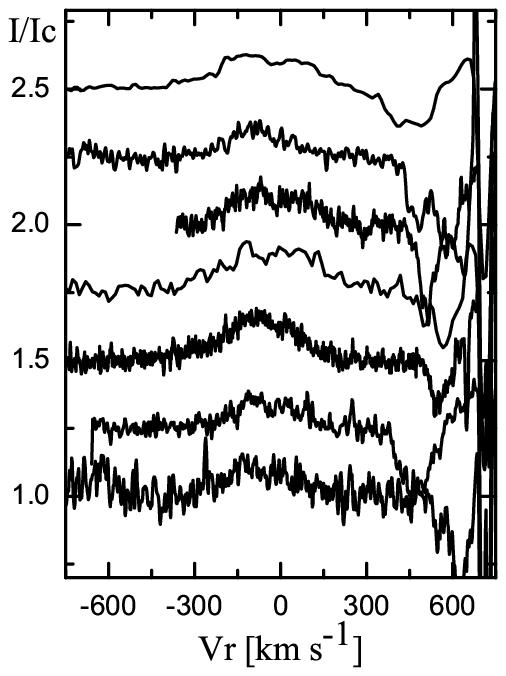}}\\
\end{tabular}
\caption{Variations of the absorption component of the H$\alpha$
line (the left panel) vs. variations of the He {\sc I} 5876 \AA\
line and the Na {\sc I} 5889 \AA\ line (the right panel). The
spectra used (from bottom to top): 2003 August 14, 2004 March 8,
2004 October 6, 2005 November 11, 2005 December 18, 2006 December
26--27 (averaged), and 2007 November 12--16 (averaged). Intensities
and heliocentric radial velocities are in the same units as in
Figure \ref{f3}. The radial velocity in the right panel is shown
with respect to the He {\sc I} line position. \label{f11}}
\end{figure*}

The profile shapes vary with time (see Figure \ref{f11}). The
position of the emission component in the P\,Cyg type profiles
depends on the strength of the absorption component. However, it
seems that the entire spectrum shifts with time. This is based on
the observed differences of the radial velocities of the emission
components of Fe {\sc II}, Na {\sc I}, and Balmer lines measured in
different spectra. The differences have the same sign and are close
in magnitude.

The shape of the absorption components in the P\,Cyg type profiles
is complicated and variable. This is clearly seen in the strongest
lines, such as the Balmer lines (see Figure \ref{f11}), Na {\sc I}
and Fe {\sc II} (mult. 42). These lines have multiple absorption
components, whose boundaries (especially the blue ones) are very
sharp. In particular, strong absorption components in the spectrum
of 2005 November 11 are almost rectangular. Their depth is nearly
the same in the range of radial velocities from $-$200 to $-$100
km\,s$^{-1}$ in the Fe {\sc II} (mult. 42) lines and from $-$250 to
$-$90 km\,s$^{-1}$ in H$\beta$ and H$\gamma$.

Our spectra obtained in 2005 November with the 6-m telescope of SAO
RAS show that the profile shape varies noticeably in just a few
days. This is detected in the Fe {\sc II} 5316 \AA\ and 5363 \AA\
lines. The line intensities did not change between 11 and 14
November, but their emission peaks shifted by +8 km\,s$^{-1}$, while
the deepest depressions in their absorption components got shifted
by $-$20 km\,s$^{-1}$. On the other hand, in our short series of
three spectra taken at San Pedro Martir in 2006 December 2006 and
2007 November (which also have a lower resolution) such shifts are
within the measurement uncertainties.

\begin{figure}
\figurenum{12} \resizebox{10cm}{!}{\includegraphics{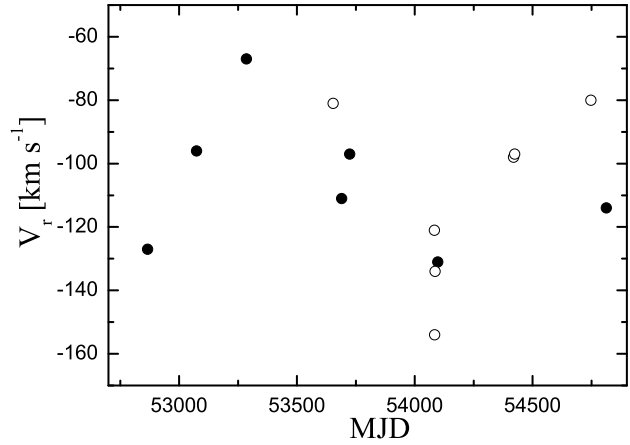}}
\caption{Variations of the Gaussian mean radial velocity of the He
{\sc I} 5876 \AA\ and 6678 \AA\ lines. The $R$=60000 data are shown
by filled circles (typical error of the velocity measurement is
$\sim$1-3 km\,s$^{-1}$), and the $R$=15000 data by open circles
(error $\sim$5--10 km\,s$^{-1}$). Radial velocities are
heliocentric, and time scale is given in modified Julian dates
(JD$-$2400000). \label{f12}}
\end{figure}

One of the most interesting findings from our spectroscopic data is
the relative behavior of the He {\sc I} 5876 \AA\ and 6678 \AA\
emission lines compared to the other observed emission lines. First,
the He {\sc I} lines show pure emission profiles that are broader
than the emission components of the Balmer and Fe {\sc II} lines.
Second, their average radial velocity is $\sim -100$ km\,s$^{-1}$
versus $\sim +20$ km\,s$^{-1}$ for the emission components of the
Balmer lines and $\sim -25$ km\,s$^{-1}$ for those of the Fe {\sc
II} lines. A similar phenomenon was found in several other FS\,CMa
type objects, such as MWC\,657 \citep{m00} and AS\,160 \citep{m03}.
Finally, they appear to exhibit a long-term ($\sim$ 1200 days),
probably cyclic, variation (Figure \ref{f12}). He {\sc I} lines
usually form near the base of the stellar wind, and the observed
variations may be due to those of the mass loss. This process is
also reflected in the variable CS components of the H$\alpha$ and Na
{\sc I} lines (see Figure \ref{f11}).

\subsection{CS dust}\label{dust}

\begin{figure}
\figurenum{13} \resizebox{10cm}{!}{\includegraphics{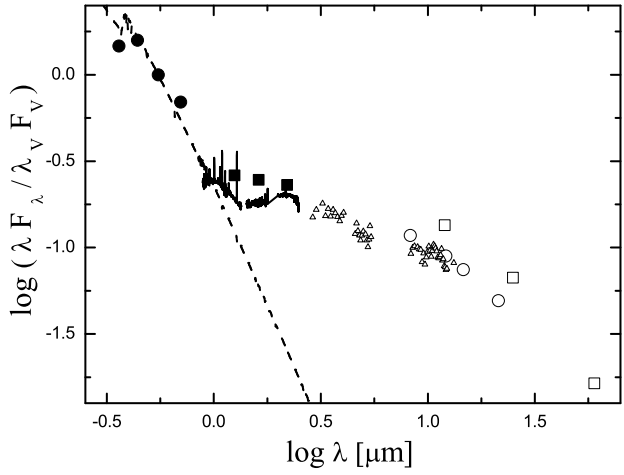}}
\caption{Spectral energy distributions of IRAS\,00470+6429.
Logarithm of the dereddened ($E(B-V)$ = 1.2 mag) flux normalized to
that in the $V$ band (vertical axis) is plotted vs. logarithm of the
wavelength in microns. The fluxes were dereddened using the
interstellar extinction law from \citet{sm79}. Symbols: circles,
$UBVR$ photometry; squares, near-IR photometry from 2MASS; the solid
line, the 2003 NIRIS data; triangles, BASS data; large open circles,
{\it MSX} data; and large open squares, IRAS data. The dashed line
represents a \citet{kur94} model atmosphere for T$_{\rm eff}$ =
20000 K. \label{f13}}
\end{figure}

The strong IR excess observed in IRAS\,00470+6429, whose SED shown
in Figure \ref{f13} is definitely due to CS dust emission. It is
clearly seen that the {\it IRAS} data obtained through large
apertures (1$\arcmin-2\arcmin$) deviate from those of BASS and {\it
MSX}. The {\it IRAS} data include a contribution from the entire H
{\sc II} region and perhaps additional emission from interstellar IR
cirrus at $\lambda$ = 60 $\mu$m (see Section \ref{evolution}).

The SED does not show any noticeable silicate feature at $\lambda
9.7 \mu$m. This might be due to a large average grain size, the
absence of silicates in the CS dust, or a large optical depth. At
this point, it is unclear which of these explanations is correct. It
seems unlikely that the grains are large, because the entire
evolutionary phase of FS\,CMa objects does not seem to last long.
Otherwise, we would observe many more such objects, because they
populate a large region of the Hertzsprung-Russell diagram.

The dust chemical composition is still uncertain. Our observations
of FS\,CMa objects with the InfraRed Spectrograph (IRS) of the {\it
Spitzer Space Telescope} show that a significant fraction of them
exhibit weak silicate emission features \citep{m08}. Unfortunately,
IRAS\,00470+6429 was not observed in this program because of the
presence of a bright nearby IR source. However, our BASS spectrum
seems to rule out silicate dust with a low optical depth, which
would produce a strong emission at $\lambda 9.7 \mu$m.

\section{Discussion}\label{discuss}

Let us consider the nature and evolutionary state of
IRAS\,00470+6429 in more detail. As mentioned in Section
\ref{intro}, the object has only been suggested to be a PPN
candidate \citep{meix99,kh05}. \citet{m07} discussed general
differences between the FS\,CMa objects and post-AGB stars. Here we
attempt a deeper analysis for IRAS\,00470+6429.

\subsection{Evolutionary state}\label{evolution}

\citet{meix99} were the first to select the object as a PPN
candidate, but they have not justified their choice. It might be due
to the 100--$\mu$m IRAS flux \citep[$10.5\pm1.2$ Jy,][]{iras86},
which is stronger than those at 12, 25, and 60 $\mu$m. However, as
IRAS\,00470+6429 is located near the Galactic plane ($b = 1\fdg9$),
its IR fluxes may be affected by the interstellar dust emission (IR
cirrus) and contaminated by emission from projectionally close
sources. Co-addition of all IRAS scans through the source position
\citep[the SCANPI procedure,][]{wj92} shows that the spatially wide
100--$\mu$m IRAS band at least partially include two optically
brighter nearby late-type stars; \object{BD$+63\arcdeg98$} at
2$\arcmin$ and \object{V634\,Cas} at 4$\arcmin$ from the
IRAS\,00470+6429. Another contribution may come from the H {\sc II}
region Sh2--182, which is considered to be illuminated by the object
\citep{rag07}. Finally, the surface brightness of the sky region,
which surrounds IRAS\,00470+6429, at $\lambda100\mu$m is high enough
for the IR cirrus emission to be important. It is represented by the
{\it cirr3} parameter in the IRAS Point Source Catalog. It is of the
order of 71$\pm$3 MJy\,ster$^{-1}$ for IRAS\,00470+6429 and nearby
objects. According to \citet{ie95}, IR cirrus affect both the
60--$\mu$m (F$_{60}$) and 100--$\mu$m fluxes of a point source even
with a CS envelope, if {\it cirr3}/F$_{60} \ge$ 1--5. This ratio is
over 30 for IRAS\,00440+6429. Therefore, the object's high
100--$\mu$m flux is unreliable.

Imaging at $\lambda10\mu$m \citep{meix99} and spectroscopy near
$\lambda2.2\mu$m \citep{kh05} showed no features, specific to the
post-AGB evolution. On the other hand, these results are inconclusive.
Although neither \citet{meix99} nor \citet{kh05} discarded the object
from their PPN candidate lists, there is no compelling evidence that
it undergoes this evolutionary stage.

The main combination of features that distinguishes IRAS\,00470+6429
from most post-AGB objects includes an early spectral type, strong
emission-line spectrum and near-IR excess, and a flux decrease
longward of the 25--$\mu$m IRAS band. Typically the dust,
responsible for the IR excess, is mostly produced during the AGB
stage. A strong stellar wind, which weakens as the object enters the
post-AGB stage, efficiently moves the dust away, so that a small
fraction of hot dust survives near the star. This makes the near-IR
excess small and produces a well-known double-peaked SED with the
short-wavelength peak due to the direct star's radiation and the
long-wavelength peak due to the dust radiation \citep[see,
e.g.,][]{how97}. The long-wavelength peak typically has the highest
flux either in the 25--$\mu$m or in the 60--$\mu$m IRAS band.
Therefore, in an IRAS color-color diagram FS\,CMa objects may be
mixed with those at the post-AGB stage \citep[see][]{m07}.

Inspection of a recent catalog of PPNe by \citet{sssb07} shows that
it contains 70 objects, which have been classified as B- or A-type
stars. Forty two of them have IRAS data, but only seven satisfy both
photometric criteria for FS\,CMa objects \citep[see Figs. 1 and 2
of][]{m07a}. Five of these seven objects have spectra with weak
emission lines and many photospheric features. The remaining two
objects are \object{MWC\,939}, which is an unconfirmed PPN and a
candidate member of the FS\,CMa group \citep{m07}, and
\object{M2--56}, which was mentioned in \citet{m07} as an example of
a possible confusion between the FS\,CMa objects and
intermediate-mass PPNe. Thus, IRAS\,00470+6429 is not a typical PPN
of an early spectral type.

M2--56 is a post-AGB object surrounded by a bipolar nebula.
According to \citet{cas02}, its closest distance (2 kpc) coincides
with that we suggest for IRAS\,00470 +6429. One possibility for our
object to be a PPN is to evolve fast enough, so that its nebula is
still too small to be resolved. During the post-AGB stage, objects
keep roughly a constant luminosity, and their photospheric
temperature increases at a different rate that depends on the
object's mass. The highest-mass objects exhibit noticeable spectral
changes over a timescale of a few decades. These changes are
accompanied by a decrease of the optical brightness, as a wavelength
of the strongest photospheric radiation shifts toward UV domain. Our
spectroscopic data for IRAS\,00470+6429 show no change in the
spectral line content. Its optical brightness in 2003--2008 ($B =
12.7-13.3$ mag) is marginally lower than that in 1958, when
\citet{dp75} first recorded it as a 12.7--mag star in the
photographic region, which roughly corresponds to the modern
$B$-band of the Johnson photometric system. However, the accuracy of
the latter estimate as well as that of its transformation into the
$B$-magnitude system cannot be verified. Also, recognized high-mass
PPNe with strong emission-line spectra that also show evidence for a
rapid spectral evolution \citep[see, e.g.,
\object{OY\,Gem},][]{jaj96,arkh06} exhibit much stronger forbidden
lines and virtually no near-IR excess. Finally, our luminosity
estimate for the object is at least a factor of two lower than those
expected for a fast-evolving post-AGB object with physical parameter
changes to be detected over a few decades \citep{bl95}.

IRAS\,00470+6429 also has some similarities with lower-mass post-AGB
objects. These objects (RV Tau stars) pulsate with periods of
30--150 days due to being located within an instability strip. A
group of them, known as RVb type, shows additional long-term
photometric variations with periods from 600 to 1500 days
\citep{dr06}. Some RV Tau stars are binary systems surrounded by
dusty disks. There is a marginal detection of variability with a
timescale of $\sim$1200 days in the photometry (see Figure
\ref{f5}). There is also independent, spectroscopic evidence (see
Section \ref{binary}) that the system is a binary with a
gaseous-and-dusty disk. On the other hand, RV Tau stars exhibit
spectroscopic properties of supergiants with weak emission-line
spectra. They also have mostly F--K spectral types and a significant
deficit of iron \citep{dr06}. If IRAS\,00470+6429 is an extremely
evolved RV Tau object, then one could expect to see some features of
a planetary nebula, such as an expanding shell similar to the one
observed in M2--56, and evidence of a chemical evolution in the
spectrum. No observation reported in this paper clearly supports
that IRAS\,00470+6429 belongs to the RV Tau group.

\subsection{Single or binary?}\label{binary}

FS\,CMa objects exhibit some features that are indicative of the
presence of secondary stellar companions. Some of them show
absorption lines that are typical for late-type stars (e.g., Li {\sc
I} 6708 \AA\ and Ca {\sc I} 6717 \AA) in addition to emission lines
that require the presence of a hot star. These are \object{MWC\,623}
\citep{z01}, \object{MWC\,728}, \object{AS\,174}, and
\object{FX\,Vel} \citep{m07a}. Also, \object{CI\,Cam} has a
degenerate secondary (a white dwarf or a neutron star), which is
thought to be responsible for a major eruption in the system in 1998
\citep[see, e.g.,][]{clark00}. A traveling emission line of He {\sc
II} at 4686 \AA\ with a similar profile to that of the He {\sc I}
lines in IRAS\,00470+6429 was found in the spectrum of CI\,Cam by
\citet{bars06}.

The absence of high-excitation lines and the presence of a weak
absorption line of Li {\sc I} at 6708 \AA\ in the spectrum of
IRAS\,00470+6429 suggest that its secondary companion is a late-type
object. The line was detected in all our spectra taken at McDonald
and has an equivalent width of 0.04 \AA. Its heliocentric radial
velocity was $-$4 km\,s$^{-1}$ in 2005 December, $-$14 km\,s$^{-1}$
in 2006 December, and $-$18 km\,s$^{-1}$ in 2008 December. This
virtual constance is inconsistent with the strong variations of the
He {\sc I} lines, so that these phenomena may not be related to each
other. This information is insufficient to constrain the putative
secondary's properties.

The presence of the Li {\sc I} line might also be due to convection
as a result of the Hot Bottom Burning (HBB) process, which occurs
during the AGB stage \citep{mac02}. However, for this process to
occur, the star has to be at a high end of the mass range of stars
that follow this evolutionary path. If IRAS\,00470+6429 is a
post-AGB object, its luminosity we derive here is not high enough to
have HBB. \citet{mac02} observed several objects in this luminosity
range (for example, NGC\,1866 $\#6, \#7, \#8$, and $\#9$) and have
detected no traces of the Li {\sc I} 6708 \AA\ line in their
spectra. Also, the radial velocity of this line in the spectrum of
IRAS\,00470+6429 is significantly different from that of the few
reliably detected photospheric absorption lines (e.g., Mg {\sc II}
4481 \AA) and from those of Fe {\sc II} emission lines, which are
believed to trace the star's velocity.

Using our photometric data and the fact that the Li {\sc I} line is
the only one that does not seem to belong to the hot companion, we
can roughly estimate an upper limit for the secondary's luminosity.
It has to be a factor of a few less luminous than the hot companion
near the Li {\sc I} line wavelength, otherwise traces of numerous
absorption lines of neutral metals \citep[e.g., Fe {\sc I}, Ca {\sc
I} as, for instance, seen in the spectrum of MWC\,623;][]{z01} would
appear in the spectrum. The red spectral region seems to be slightly
affected by the CS contribution, which is not well-constrained (see
Section \ref{stel_par}), but the secondary's continuum is not
noticeable. Therefore, the secondary should be at least 1 mag
fainter in the $V$-band. If it is an early K--type star, as in the
case of MWC\,623 \citep{z01}, its luminosity does not exceed
$\sim$500 L$_{\odot}$. Given the available evidence, it is not
possible to further constrain the secondary's spectral type.
Formally, this estimate assumes that the luminosity type is between
{\sc IV} and {\sc III} \citep{sk81}. A dwarf of any appropriate late
spectral type would be too dim to reveal any of its features, while
a bright giant or a supergiant would dominate in a region between
$\sim$0.5--1.5 $\mu$m (CS dust takes over at longer wavelengths).

\section{Conclusions}\label{concl}

We accomplished a five-year long (2003--2008) photometric and
spectroscopic monitoring of the optical counterpart of the IR source
IRAS\,00470+6429. Our results in combination with other available
data for the object and information about surrounding stars and
associations allowed us to reach the following conclusions.

\begin{itemize}
\item The optical counterpart of IRAS\,00470+6429 is most likely a
binary system with a B-type visible primary companion and a much
fainter late-type secondary companion. The main indication in favor
of the binary hypothesis is the presence of a weak Li {\sc I} 6708
\AA\ absorption line in the object's spectrum.
\item The primary companion is a B2--B3 spectral type with
a luminosity of $\log$ L/L$\odot$ = 3.9$\pm$0.3 at a distance of
2.0$\pm$0.3 kpc from the Sun. The system probably belongs to the
stellar association Cas\,OB7. Properties of the association members
need to be better constrained with high-resolution spectroscopy to
improve its distance and investigate the fine internal structure,
revealed in this study.
\item The optical brightness variations exceed 0.5 mag that is
typical for FS\,CMa type objects. Many lines in the spectrum of
the B-type companion exhibit variable P\,Cyg type profiles.
Also, multiple variable CS absorption components of the Na {\sc I}
D-lines were detected. This indicates the presence of
a significant amount of CS gas in the line of sight. Nevertheless,
most of the observed reddening has interstellar origin.
\item Although some similarities with post-AGB type objects can be
identified in the properties of IRAS\,00470+6429 (e.g., possible
long-term variation cycles), it is unlikely to be highly evolved.
\end{itemize}

The results of this study support our earlier hypothesis that
FS\,CMa type objects are binary systems. The sparseness of our
spectroscopic data did not allow us to find the orbital period of
the IRAS\,00470+6429 system. In the next paper of this series we
will present an initial attempt to model the wealth of data
available for this object.

\acknowledgements We thank the anonymous referee for suggestions
that helped us to improve the material presentation. The research
was partially supported by a Civilian Research and Development
Foundation (CRDF) through the grant RUP1--2687--NA--05 to K.S.B.,
V.G.K., A.M., and E.L.C., by a Russian Foundation for Basic Research
grant 08--02--0072 to V.G.K., and by a UNCG Summer Research
Excellence Grant 10060 to A.M. Also it was partially funded by the
Division of Physical Sciences of the Russian Academy of Sciences
through the program Extended objects in the Universe. This research
has made use of the SIMBAD database operated at CDS, Strasbourg,
France, the NSVS database, and data products from the Two Micron All
Sky Survey (2MASS), which is a joint project of the University of
Massachusetts and the Infrared Processing and Analysis
Center/California Institute of Technology, funded by the National
Aeronautics and Space Administration and the National Science
Foundation.

\begin{table}[H]
\caption{Observing log of the optical spectroscopic observations of
IRAS\,00470+6429}\label{t1} {\tiny
\begin{tabular}{cccccccccc}
\tableline\noalign{\smallskip}
Date      & HJD     & Range & Exp.time &Ref.& Date     & HJD    & Range & Exp.time &Ref.\\
mm/dd/yyyy& 2450000+& \AA      &   s      &    &mm/dd/yyyy&2450000+& \AA      &   s      &    \\
\noalign{\smallskip}\tableline\noalign{\smallskip}
08/14/2003& 2866.42 & 5278--6760 & 3600& 1&12/27/2006& 4096.65 & 3600--10500& 3600& 6\\
08/18/2003& 2869.81 & 3800--5600 & 3170& 2&12/28/2006& 4097.64 & 3600--10500& 3600& 6\\
03/09/2004& 3073.52 & 5278--6760 & 3600& 1&11/14/2007& 4418.75 & 3800--6800 & 4800& 5\\
10/06/2004& 3285.50 & 5278--6760 & 3600& 1&11/15/2007& 4419.76 & 3800--6800 & 4800& 5\\
11/27/2004& 3336.40 & 3600--8700 & 3600& 3&11/19/2007& 4423.76 & 3800--6800 & 4800& 5\\
12/26/2004& 3365.62 & 5400--6700 & 2700& 4&10/04/2008& 4743.88 & 3800--6800 & 3600& 5\\
10/09/2005& 3653.80 & 4300--6800 & 3600& 5&10/05/2008& 4744.78 & 3800--6800 & 3600& 5\\
10/12/2005& 3655.88 & 4300--6800 & 3600& 5&10/07/2008& 4746.77 & 3800--6800 & 6000& 5\\
11/11/2005& 3686.38 & 4015--5460 & 3600& 1&10/09/2008& 4748.79 & 3800--6800 & 4800& 5\\
11/14/2005& 3689.34 & 5278--6760 & 3600& 1&10/10/2008& 4749.82 & 3800--6800 & 4800& 5\\
12/19/2005& 3723.60 & 3600--10500& 6000& 6&10/11/2008& 4750.74 & 3800--6800 & 4800& 5\\
12/13/2006& 4082.72 & 3800--6800 & 3600& 5&12/13/2008& 4812.64 & 3600--10500& 4800& 6\\
12/14/2006& 4083.70 & 3800--6800 & 4800& 5&12/14/2008& 4813.67 & 3600--10500& 3600& 6\\
12/16/2006& 4085.73 & 3800--6800 & 3600& 5&          &         &            &     &  \\
\noalign{\smallskip}\tableline
\end{tabular}
} \tablecomments{Columns 1 and 2: calendar date and Heliocentric
Julian Date of the observation; Columns 3 and 4: spectral range and
exposure time of the spectrum; Column 5: telescopes used with
references to the spectrographs: (1) BTA RAS, the 6--m telescopes of
the Russian Academy of Sciences, Russia, NES spectrograph, $R$ =
60000, Panchuk et al. (2007); (2) 0.8--m telescope of the Dark Sky
Observatory, Boone, NC, USA, $R$ = 1300; (3) 1.52--m telescope of
the Loiano Observatory, Bologna, Italy, $R$ = 3000; (4) 2.1--m Otto
Struve telescope of the Mcdonald Observatory, Ft. Davis, TX, USA,
$R$ = 60000, McCarthy et al. (1993); (5) 2.1--m telescope of the
SPM, San Pedro Martir Observatory, Mexico, $R$ = 15000, Levine \&
Chakrabarty (1995); (6) 2.7--m Harlan J. Smith telescope of the
Mcdonald Observatory, Ft. Davis, TX, USA, $R$ = 60000, Tull et al.
(1995)}
\end{table}

\clearpage
\begin{table}[H]
\caption{Line identification, intensities and radial velocities
$V_r$ in the optical spectrum of IRAS\,00470+6429}\label{t2} {\tiny
\begin{tabular}{lccccccccc}
\noalign{\smallskip}\tableline\noalign{\smallskip}
Line     & $\lambda_{\rm lab}$& \multicolumn{2}{c}{08/14/2003} & \multicolumn{2}{c}{03/09/2004} & \multicolumn{2}{c}{10/06/2004} & \multicolumn{2}{c}{11/11/2005} \\
\cline{3-4}\cline{5-6}\cline{7-8}\cline{9-10}
         &  \AA               &   $r$      &    $V_r$          &   $r$      &    $V_r$          &   $r$      &    $V_r$          &   $r$      &    $V_r$          \\
\noalign{\smallskip}\tableline\noalign{\smallskip}
H$\delta$& 4101.74&         &          &          &         &         &           & 0.18/1.44 & -140:/45:  \\
TiII(105)& 4171.90&         &          &          &         &         &           & 0.90/1.08 &            \\
FeII(27) & 4173.46&         &          &          &         &         &           &           &            \\
FeII(21) & 4177.68&         &          &          &         &         &           & 0.87/1.10 &            \\
FeII(28) & 4178.85&         &          &          &         &         &           &           &            \\
FeII(27) & 4233.17&         &          &          &         &         &           & 0.76/1.50 &  -150/-23: \\
ScII(7)  & 4246.82&         &          &          &         &         &           &           &            \\
TiII(41) & 4290.21&         &          &          &         &         &           & 0.85/1.08 &  -162:/-46:\\
TiII(20) & 4294.10&         &          &          &         &         &           & 0.88/1.10 &  -150:/ -  \\
FeII(98) & 4296.57&         &          &          &         &         &           & 0.92/1.05 &  -140:/-58:\\
TiII(41) & 4300.04&         &          &          &         &         &           & 0.85/1.18 &  -150/-50  \\
FeII(27) & 4303.17&         &          &          &         &         &           &  -  /1.16 &    - /-54  \\
TiII(41) & 4307.89&         &          &          &         &         &           & 0.90/1.18 & -165:/-60: \\
TiII(41) & 4312.86&         &          &          &         &         &           & 0.82/ -   & -150:/ -   \\
ScII(15) & 4314.08&         &          &          &         &         &           &   - /1.10 &    - /-40: \\
FeII(32) & 4314.30&         &          &          &         &         &           &           &            \\
H$\gamma$& 4340.47&         &          &          &         &         &           & 0.08/2.08 & -162/33    \\
FeII(27) & 4351.77&         &          &          &         &         &           & 0.83/1.35 & -160/-45   \\
TiII(104)& 4367.65&         &          &          &         &         &           & 0.90/1.03 & -120:/-34: \\
FeII(27) & 4385.38&         &          &          &         &         &           &   - /1.20 &   -  /-49: \\
TiII(19) & 4395.03&         &          &          &         &         &           & 0.82/1.25 & -160/-43   \\
TiII(51) & 4399.77&         &          &          &         &         &           & 0.94/1.14 & -145:/-40: \\
FeII(27) & 4416.82&         &          &          &         &         &           & 0.90/1.20 &            \\
TiII(40) & 4417.72&         &          &          &         &         &           &           &            \\
TiII(51) & 4418.33&         &          &          &         &         &           &           &            \\
TiII(19) & 4443.80&         &          &          &         &         &           & 0.90/1.22 & -162:/-40  \\
TiII(31) & 4468.49&         &          &          &         &         &           & 0.92/1.32 & -160/-41   \\
FeII(37) & 4472.92&         &          &          &         &         &           & 0.87/1.09 & -130:/-54: \\
MgII(4)  & 4481.22&         &          &          &         &         &           & 0.84      &     -35    \\
FeII(37) & 4489.17&         &          &          &         &         &           & 0.92/1.08 & -144:/-70: \\
FeII(37) & 4491.40&         &          &          &         &         &           &   - /1.10 &   -  /-60: \\
TiII(31) & 4501.27&         &          &          &         &         &           & 0.90/1.17 & -160:/-50  \\
FeII(38) & 4508.28&         &          &          &         &         &           & 0.90/1.30 & -158/-41   \\
FeII(37) & 4515.33&         &          &          &         &         &           & 0.90/1.27 & -157/-43   \\
FeII(37) & 4520.22&         &          &          &         &         &           & 0.92/1.18 & -159:/-53: \\
FeII(38) & 4522.63&         &          &          &         &         &           & 0.87/1.33 &   -  /-41  \\
TiII(82) & 4529.48&         &          &          &         &         &           & 0.92/1.03 &            \\
TiII(50) & 4534.02&         &          &          &         &         &           & 0.84/1.25 & -165/-47:  \\
FeII(37) &        &         &          &          &         &         &           &           &            \\
FeII(38) & 4541.51&         &          &          &         &         &           & 0.91/1.10 & -161:/-43: \\
FeII(38) & 4549.54&         &          &          &         &         &           & 0.68/1.55 & -160/-34   \\
TiII(82) &        &         &          &          &         &         &           &           &            \\
FeII(37) & 4555.89&         &          &          &         &         &           & 0.93/1.31 & -151:/-47: \\
CrII(44) & 4558.65&         &          &          &         &         &           &           &            \\
TiII(50) & 4563.76&         &          &          &         &         &           & 0.89/1.17 & -161/-51   \\
TiII(82) & 4571.97&         &          &          &         &         &           & 0.83/1.24 & -168:/-40  \\
FeII(38) & 4576.33&         &          &          &         &         &           & 0.90/1.03 &   -  /-63: \\
FeII(37) & 4582.83&         &          &          &         &         &           &           &            \\
FeII(38) & 4583.83&         &          &          &         &         &           & 0.74/1.60 & -157/-36   \\
CrII(44) & 4588.20&         &          &          &         &         &           & 0.90/1.09 &    - /-58: \\
CrII(44) & 4589.89&         &          &          &         &         &           &           &            \\
TiII(50) & 4589.95&         &          &          &         &         &           &           &            \\
CrII(44) & 4618.82&         &          &          &         &         &           & 0.96/1.10 &            \\
FeII(38) & 4620.51&         &          &          &         &         &           &           &            \\
FeII(37) & 4629.33&         &          &          &         &         &           & 0.93/1.36 & -154/-43   \\
CrII(44) & 4634.07&         &          &          &         &         &           & 0.95/1.14 &            \\
FeII(186)& 4635.31&         &          &          &         &         &           &           &            \\
DIB      & 4726.27&         &          &          &         &         &           &           &            \\
FeII(43) & 4731.47&         &          &          &         &         &           & 0.94/1.06 & -165:/-27: \\
DIB      & 4762.67&         &          &          &         &         &           &           &            \\
TiII(17) & 4762.78&         &          &          &         &         &           &           &            \\
TiII(48) & 4763.89&         &          &          &         &         &           & 0.91/ -   &            \\
TiII(48) & 4764.53&         &          &          &         &         &           &           &            \\
CrII(30) & 4824.14&         &          &          &         &         &           & 0.93/1.09 & -157:/-55: \\
CrII(30) & 4848.25&         &          &          &         &         &           & 0.94/1.12 & -165:/-55: \\
H$\beta$ & 4861.33&         &          &          &         &         &           & 0.90:     &  -380:     \\
         &        &         &          &          &         &         &           & 0.17/3.90 &  -175/36   \\
TiII(114)& 4911.19&         &          &          &         &         &           & 0.91/ -   &            \\
FeII(42) & 4923.92&         &          &          &         &         &           & 0.86      &  -240      \\
         &        &         &          &          &         &         &           & 0.52/1.86 &  -155/-26  \\
OI(14)   & 4967.38&         &          &          &         &         &           &           &            \\
OI(14)   & 4967.88&         &          &          &         &         &           & 0.94      &            \\
OI(14)   & 4968.79&         &          &          &         &         &           &           &            \\
FeII(42) & 5018.44&         &          &          &         &         &           & 0.76      &  -235      \\
         &        &         &          &          &         &         &           & 0.43/2.02 &  -158/-23  \\
SiII(5)  & 5056.02&         &          &          &         &         &           &           &            \\
FeII(167)& 5127.86&         &          &          &         &         &           &           &            \\
TiII(86) & 5129.16&         &          &          &         &         &           & 0.96/1.02 & -170:/-45: \\
\noalign{\smallskip}\tableline\
\end{tabular}
} \addtocounter{table}{-1}
\end{table}

\begin{table}[H]
\caption{Line identification, intensities and radial velocities
$V_r$ in the optical spectrum of IRAS\,00470+6429. Continued} {\tiny
\begin{tabular}{lccccccccc}
\noalign{\smallskip}\tableline\noalign{\smallskip}
Line     & $\lambda_{\rm lab}$& \multicolumn{2}{c}{08/14/2003} & \multicolumn{2}{c}{03/09/2004} & \multicolumn{2}{c}{10/06/2004} & \multicolumn{2}{c}{11/11/2005} \\
\cline{3-4}\cline{5-6}\cline{7-8}\cline{9-10}
         &  \AA               &   $r$      &    $V_r$          &   $r$      &    $V_r$          &   $r$      &    $V_r$          &   $r$      &    $V_r$          \\
\noalign{\smallskip}\tableline\noalign{\smallskip}
FeII(35) & 5132.66&         &          &          &         &         &           & 0.96/1.03 &   -  /-49: \\
FeII(35) & 5146.11&         &          &          &         &         &           & 0.95/1.02 &            \\
TiII(70) & 5154.07&         &          &          &         &         &           & 0.95/1.03 &  -161/-60: \\
FeII(42) & 5169.03&         &          &          &         &         &           & 0.62      &  -235      \\
         &        &         &          &          &         &         &           & 0.35/1.97 &  -156/-21  \\
MgI(2)   & 5172.69&         &          &          &         &         &           &           &            \\
MgI(2)   & 5183.61&         &          &          &         &         &           & 0.95/1.16 & -170:/-54: \\
TiII(70) & 5188.68&         &          &          &         &         &           & 0.97/1.13 & -140:/-60: \\
FeII(49) & 5197.58&         &          &          &         &         &           & 0.95/1.38 & -162:/-50: \\
TiII(70) & 5226.54&         &          &          &         &         &           & 0.95/1.12 &   -  /-50: \\
FeII(49) & 5234.62&         &          &          &         &         &           & 0.94/1.37 &  -147/-46  \\
FeII(49) & 5254.93&         &          &          &         &         &           & 0.96/1.02 &            \\
FeII(48) & 5264.80&         &          &          &         &         &           & 0.95/-    & -146:/ -   \\
TiII(103)& 5268.63&         &          &          &         &         &           & 0.95/1.02 &            \\
FeII(185)& 5272.40&         &          &          &         &         &           & 0.93/1.02 & -135:/-50: \\
FeII(49) & 5276.00&         &          &          &         &         &           & 0.92/1.38 & -174:/-44  \\
CrII(43) & 5279.88&         &          &          &         &         &           & 0.92/ -   &            \\
FeII(40) & 5284.10& - /1.24 & - /-10:  &          &         &         &           & 0.92/1.13 &  -160/-37  \\
FeII(49) & 5316.65&0.90/1.70&-150:/-16 &0.87/1.50:&-272/-42 &0.96/1.85& -200/-21  & 0.90/1.69 &  -160/-41  \\
FeII(49) & 5325.56&         &          &          &         &         &           & 0.95/ -   &            \\
OI(12)   & 5329.10&         &          &          &         &         &           &           &            \\
OI(12)   & 5329.69&         &          &          &         &         &           & 0.91      &            \\
OI(12)   & 5330.74&         &          &          &         &         &           &           &            \\
FeII(48) & 5362.86&0.96/1.23&-170:/-28 &0.97:/1.16&-270/-40 &0.98/1.19& -193:/-21:& 0.95/1.21 & -157:/-47  \\
DIB      & 5404.50&         &          &          &         &0.92     &    -8:    & 0.95      &            \\
FeII(49) & 5425.25&         &          &          &         &0.98/1.06&  -222/ -  &           &            \\
FeII(55) & 5432.98&         &          &          &         &         &           &           &            \\
DIB      & 5487.67&         &          &          &         &         &           &           &            \\
DIB      & 5494.10&         &          &          &         &         &           &           &            \\
FeII(55) & 5534.84&         &          &0.97:1.09 &-165:/-51&0.97/1.12& -195/-22: &           &            \\
DIB      & 5762.70&         &          &          &         &         &           &           &            \\
DIB      & 5766.16&         &          &          &         &         &           &           &            \\
DIB      & 5780.37&  0.76   & -7:      &0.77:     &-14:     &0.71     &    -8:    &           &            \\
DIB      & 5796.96&  0.79   & -9:      &0.80      &-13      &0.80     &   -14     &           &            \\
DIB      & 5809.24&         &          &          &         &         &           &           &            \\
DIB      & 5849.80&         &          &0.94:     &-10:     &0.88     &    -9:    &           &            \\
HeI(11)  & 5875.72&  1.12   & -127:    &1.14:     &-67:     &1.17     &   -96     &           &            \\
NaI(1)   & 5889.95&  0.80   & -180     &0.71/1.12:&-263/-34 &0.80     &  -190     &           &            \\
         &        &0.60/1.30&-108/25:  &          &         &0.91/1.27& -135/0    &           &            \\
         &        &  0.04   & -17      &0.06      &-18      &0.03     &   -18     &           &            \\
NaI(1)   & 5895.92&  0.92   & -175     &0.87/1.15:&-225/-9: &0.92     &  -178     &           &            \\
         &        &0.73/1.22&-102/18:  &          &         &1.01/1.30& -115:/3   &           &            \\
         &        &  0.03   &  -17     &0.06      &-18      &0.04     &   -17     &           &            \\
FeII(46) & 5991.37&         &          &          &         &         &           &           &            \\
FeII(74) & 6147.74& - /1.17 &          &          &         &- /1.12  &           &           &            \\
FeII(74) & 6149.25&         &          &          &         &         &           &           &            \\
DIB      & 6195.96&         &          &0.87      &-12      &0.86     &   -15     &           &            \\
DIB      & 6203.08&         &          &0.88      &-20:     &0.91     &   -16     &           &            \\
FeII(74) & 6238.39&         &          &          &         &         &           &           &            \\
FeII(74) & 6239.95&         &          &          &         &         &           &           &            \\
FeII(74) & 6247.55&0.94/1.2 & -178:/-6 &          &         &- /1.17  &    - /-14:&           &            \\
DIB      & 6269.75&   0.86  & -14:     &0.88      &-8:      &0.89     &   -11:    &           &            \\
DIB      & 6283.85&   0.79  & -12:     &0.78      &-11:     &0.82     &   -14:    &           &            \\
$[OI]1F$ & 6300.30&   1.40  & -42:     &1.28      &-38      &1.30     &   -37:    &           &            \\
$[OI]1F$ & 6363.78&   1.13  & -37:     &1.08      &-41      &1.10     &   -39     &           &            \\
DIB      & 6375.95&         &          &0.92:     &-12:     &0.86     &   -12     &           &            \\
DIB      & 6379.29&   0.77  & -16:     &0.79:     &-13      &0.79     &   -18     &           &            \\
FeII(74) & 6416.93&         &          &          &         &0.96/1.05& -143:/14  &           &            \\
FeII(40) & 6432.70&0.96/1.0 &-130:/-20:&-/1.07    &-/-38:   &0.97/1.07& -155/-14  &           &            \\
FeII(74) & 6456.38&0.96/1.2 & -100:/-3 &0.97/1.15 &-226/-30 &-  /1.30 &    - /-18 &           &            \\
FeII(40) & 6516.08& - /1.2  & -  /-15: &-/1.08:   &-/-36:   &- /1.19  &    - /-25 &           &            \\
H$\alpha$& 6562.81&0.53     & -310     &0.51/11.9 &-320/-13 &0.60     &  -540:    &           &            \\
         &        &0.53/16. & -220:/67 &          &         &0.46/15.2& -377/18   &           &            \\
DIB      & 6613.56&   0.75  &    -16   &0.77      &-13      &0.78     &    -15    &           &            \\
DIB      & 6660.64&         &          &0.91      &-11:     &         &           &           &            \\
HeI(46)  & 6678.15&         &          &          &         &         &           &           &            \\
LiI(1)   & 6707.80&         &          &          &         &         &           &           &            \\
FeII(73) & 7462.38&         &          &          &         &         &           &           &            \\
Pa 29    & 8292.31&         &          &          &         &         &           &           &            \\
Pa 20    & 8392.40&         &          &          &         &         &           &           &            \\
Pa 19    & 8413.31&         &          &          &         &         &           &           &            \\
CaII(2)  & 8542.09&         &          &          &         &         &           &           &            \\
Pa 14    & 8598.39&         &          &          &         &         &           &           &            \\
DIB      & 8620.00&         &          &          &         &         &           &           &            \\
NI(7)    & 8629.24&         &          &          &         &         &           &           &            \\
Pa 12    & 8750.47&         &          &          &         &         &           &           &            \\
Pa 10    & 9014.91&         &          &          &         &         &           &           &            \\
Pa 9     & 9229.00&         &          &          &         &         &           &           &            \\
Pa  7    &10049.40&         &          &          &         &         &           &           &            \\
\noalign{\smallskip}\tableline
\end{tabular}
} \addtocounter{table}{-1}
\end{table}

\begin{table}[H]
\caption{Line identification, intensities and radial velocities
$V_r$ in the optical spectrum of IRAS\,00470+6429. Continued} {\tiny
\begin{tabular}{lccccccccc}
\noalign{\smallskip}\tableline\noalign{\smallskip}
Line     & $\lambda_{\rm lab}$& \multicolumn{2}{c}{11/14/2005} & \multicolumn{2}{c}{12/18/2005}& \multicolumn{2}{c}{12/27/2006}& \multicolumn{2}{c}{12/14/2008}\\
\cline{3-4}\cline{5-6}\cline{7-8}\cline{9-10}
         &  \AA   &   $r$     & $V_r$   &   $r$   & $V_r$   &   $r$   &  $V_r$  &    $r$   &    $V_r$   \\
\noalign{\smallskip}\tableline\noalign{\smallskip}
H$\delta$& 4101.74&          &          &0.26/1.45&$-$251/35&0.14/1.37&$-$173/54& 0.14/1.45&$-$311/19   \\
         &        &          &          &0.40     &$-$421   &         &         &          &            \\
TiII(105)& 4171.90&          &          &         &         &         &         &          &            \\
FeII(27) & 4173.46&          &          &         &         &         &         &          &            \\
FeII(21) & 4177.68&          &          &         &         &         &         &          &            \\
FeII(28) & 4178.85&          &          &         &         &         &         &          &            \\
FeII(27) & 4233.17&          &          &0.90/1.39&$-$244/$-$27& $-$/1.47& $-$/$-$24:&$-$/1.47&$-$/$-$20\\
ScII(7)  & 4246.82&          &          &  1.14   &$-$42:   &  1.15   &$-$34:   & 1.15     &$-$31       \\
TiII(41) & 4290.21&          &          &         &         &         &         &          &            \\
TiII(20) & 4294.10&          &          &         &         &         &         &          &            \\
FeII(98) & 4296.57&          &          &         &         &         &         &          &            \\
TiII(41) & 4300.04&          &          &         &         &         &         &          &            \\
FeII(27) & 4303.17&          &          &         &         &         &         &          &            \\
TiII(41) & 4307.89&          &          &         &         &         &         &          &            \\
TiII(41) & 4312.86&          &          &         &         &         &         &          &            \\
ScII(15) & 4314.08&          &          &         &         &         &         &          &            \\
FeII(32) & 4314.30&          &          &         &         &         &         &          &            \\
H$\gamma$& 4340.47&          &          &0.25/1.88&$-$237/29&0.14/1.96&$-$205/24&0.16/2.15 &$-$319/13   \\
         &        &          &          &0.37     &$-$411   &         &         &          &            \\
FeII(27) & 4351.77&          &          &$-$/1.43 &$-$/$-$37&$-$/1.42 &$-$/$-$33&$-$/1.41  &$-$/$-$41   \\
TiII(104)& 4367.65&          &          &         &         &         &         &          &            \\
FeII(27) & 4385.38&          &          &         &         &         &         &          &            \\
TiII(19) & 4395.03&          &          &         &         &$-$/1.29 &$-$/$-$30&          &            \\
TiII(51) & 4399.77&          &          &         &         &$-$/1.13 &$-$/$-$36&          &            \\
FeII(27) & 4416.82&          &          &         &         &$-$/1.20 &$-$/$-$17:&$-$/1.22 &$-$/$-$29   \\
TiII(40) & 4417.72&          &          &         &         &         &         &          &            \\
TiII(51) & 4418.33&          &          &         &         &         &         &          &            \\
TiII(19) & 4443.80&          &          &$-$/1.19 &$-$/$-$30&$-$/1.17 &$-$/$-$38&$-$/1.17  &$-$/$-$31   \\
TiII(31) & 4468.49&          &          &$-$/1.20 &$-$/$-$34&$-$/1.20 &$-$/$-$31&$-$/1.23  &$-$/$-$36   \\
FeII(37) & 4472.92&          &          &         &         &         &         &          &            \\
MgII(4)  & 4481.22&          &          &         &         &0.86     &$-$29    &0.85      &$-$28       \\
FeII(37) & 4489.17&          &          &         &         &         &         &          &            \\
FeII(37) & 4491.40&          &          &         &         &         &         &          &            \\
TiII(31) & 4501.27&          &          &         &         &         &         &          &            \\
FeII(38) & 4508.28&          &          &         &         &         &         &$-$/1.22  &$-$/$-$44   \\
FeII(37) & 4515.33&          &          &         &         &         &         &$-$/1.21  &$-$/$-$44   \\
FeII(37) & 4520.22&          &          &         &         &         &         &          &            \\
FeII(38) & 4522.63&          &          &         &         &         &         &          &            \\
TiII(82) & 4529.48&          &          &         &         &         &         &          &            \\
TiII(50) & 4534.02&          &          &         &         &         &         &$-$/1.24  &$-$/$-$38   \\
FeII(37) &        &          &          &         &         &         &         &          &            \\
FeII(38) & 4541.51&          &          &         &         &         &         &          &            \\
FeII(38) & 4549.54&          &          &         &         &         &         &          &            \\
TiII(82) &        &          &          &         &         &         &         &          &            \\
FeII(37) & 4555.89&          &          &         &         &         &         &          &            \\
CrII(44) & 4558.65&          &          &         &         &         &         &          &            \\
TiII(50) & 4563.76&          &          &         &         &         &         &          &            \\
TiII(82) & 4571.97&          &          &         &         &         &         &          &            \\
FeII(38) & 4576.33&          &          &         &         &         &         &          &            \\
FeII(37) & 4582.83&          &          &         &         &         &         &          &            \\
FeII(38) & 4583.83&          &          &         &         &         &         &          &            \\
CrII(44) & 4588.20&          &          &         &         &         &         &          &            \\
CrII(44) & 4589.89&          &          &         &         &         &         &          &            \\
TiII(50) & 4589.95&          &          &         &         &         &         &          &            \\
CrII(44) & 4618.82&          &          &         &         &         &         &          &            \\
FeII(38) & 4620.51&          &          &         &         &         &         &          &            \\
FeII(37) & 4629.33&          &          &         &         &         &         &          &            \\
CrII(44) & 4634.07&          &          &         &         &         &         &          &            \\
FeII(186)& 4635.31&          &          &         &         &         &         &          &            \\
DIB      & 4726.27&          &          &         &         &         &         &          &            \\
FeII(43) & 4731.47&          &          &         &         &         &         &          &            \\
DIB      & 4762.67&          &          &         &         &         &         &          &            \\
TiII(17) & 4762.78&          &          &         &         &         &         &          &            \\
TiII(48) & 4763.89&          &          &         &         &         &         &          &            \\
TiII(48) & 4764.53&          &          &         &         &         &         &          &            \\
CrII(30) & 4824.14&          &          &         &         &         &         & 1.11     &$-$31:      \\
CrII(30) & 4848.25&          &          &         &         &         &         &          &            \\
H$\beta$ & 4861.33&          &          &1.09     &$-$859:  &         &         &          &            \\
         &        &          &          &0.28/3.33&$-$285/39&0.23/3.30&$-$237/37&0.16/3.26 &$-$321/8    \\
TiII(114)& 4911.19&          &          &         &         &         &         &          &            \\
FeII(42) & 4923.92&          &          &0.91     &$-$440   &0.61/1.73&$-$152/$-$2&        &            \\
         &        &          &          &0.74/1.69&$-$223/$-$21&      &         &0.78/1.73 &$-$458/$-$35\\
OI(14)   & 4967.38&          &          &         &         &         &         &          &            \\
OI(14)   & 4967.88&          &          &         &         &         &         &          &            \\
OI(14)   & 4968.79&          &          &         &         &         &         &          &            \\
FeII(42) & 5018.44&          &          &         &         &         &         &          &            \\
         &        &          &          &0.77/1.98&$-$219/$-$21&0.78/1.98&$-$205/$-$18 &0.67/2.04 &$-$471/$-$35\\
\noalign{\smallskip}\tableline
\end{tabular}
} \addtocounter{table}{-1}
\end{table}

\begin{table}[H]
\caption{Line identification, intensities and radial velocities
$V_r$ in the optical spectrum of IRAS\,00470+6429. Continued} {\tiny
\begin{tabular}{lccccccccc}
\noalign{\smallskip}\tableline\noalign{\smallskip}
Line     & $\lambda_{\rm lab}$& \multicolumn{2}{c}{11/14/2005} & \multicolumn{2}{c}{12/18/2005} & \multicolumn{2}{c}{12/27/2006}&\multicolumn{2}{c}{12/14/2008}\\
\cline{3-4}\cline{5-6}\cline{7-8}\cline{9-10}
         &  \AA              &   $r$    &    $V_r$          &   $r$   &    $V_r$ &   $r$   &    $V_r$ \\
\noalign{\smallskip}\tableline\noalign{\smallskip}
SiII(5)  & 5056.02&          &          &0.95     &$-$22    &         &          &0.95     &$-$23     \\
FeII(167)& 5127.86&          &          &         &         &         &          &         &          \\
TiII(86) & 5129.16&          &          &         &         &         &          &         &          \\
FeII(35) & 5132.66&          &          &         &         &         &          &         &          \\
FeII(35) & 5146.11&          &          &         &         &         &          &         &          \\
TiII(70) & 5154.07&          &          &         &         &         &          &         &          \\
FeII(42) & 5169.03&          &          &         &         &0.83     &$-$360    &0.90:    &$-$461    \\
         &        &          &          &0.75/1.95&$-$230/$-$27&0.67/1.85&$-$155/$-$16&$-$/2.03&$-$/$-$35\\
MgI(2)   & 5172.69&          &          &         &         &1.16     &$-$35     &         &          \\
MgI(2)   & 5183.61&          &          &$-$/1.16 &$-$/$-$36&$-$/1.13 &$-$/$-$30 &$-$/1.21 &$-$/$-$28 \\
TiII(70) & 5188.68&          &          &         &         &$-$/1.06 &$-$/$-$41 &$-$/1.07 &$-$/$-$55:\\
FeII(49) & 5197.58&          &          &$-$/1.33 &$-$/$-$38&0.95/1.34&$-$327:/$-$34&0.96/1.31&$-$310/$-$29\\
TiII(70) & 5226.54&          &          &$-$/1.08 &$-$/$-$37&         &          &         &          \\
FeII(49) & 5234.62&          &          &$-$/1.39 &$-$/$-$38&         &          &         &          \\
FeII(49) & 5254.93&          &          &         &         &         &          &         &          \\
FeII(48) & 5264.80&          &          &         &         &         &          &         &          \\
TiII(103)& 5268.63&          &          &         &         &         &          &         &          \\
FeII(185)& 5272.40&          &          &         &         &         &          &         &          \\
FeII(49) & 5276.00&          &          &0.95/1.36&$-$226/$-$42&$-$/1.36&$-$/$-$35&$-$/1.35&$-$/$-$39 \\
CrII(43) & 5279.88&          &          &         &         &         &          &         &          \\
FeII(40) & 5284.10&          &          &$-$/1.10 &$-$/$-$38&$-$/1.08 &$-$/$-$37 &$-$/1.11 &$-$/$-$38 \\
FeII(49) & 5316.65&0.93/1.64 &$-$191/$-$33&0.94/1.61&$-$235/$-$39&$-$/1.74&$-$/$-$34&$-$/1.54&$-$/$-$41\\
FeII(49) & 5325.56&          &          &         &         &         &          &         &          \\
OI(12)   & 5329.10&          &          &         &         &         &          &         &          \\
OI(12)   & 5329.69&          &          &         &         &         &          &         &          \\
OI(12)   & 5330.74&          &          &         &         &         &          &         &          \\
FeII(48) & 5362.86&0.97/1.21 &$-$206:/$-$35&$-$/1.18&$-$/$-$36&$-$/1.19&$-$/$-$36&$-$/1.20 &$-$/$-$42 \\
DIB      & 5404.50&   0.94   &$-$10:    &         &         &         &          &         &          \\
FeII(49) & 5425.25&$-$/1.04  &$-$/$-$33 &         &         &         &          &         &          \\
FeII(55) & 5432.98&0.98/1.02 &$-$132:/$-$40:&     &         &         &          &         &          \\
DIB      & 5487.67&    0.95  &$-$10:    &         &         &         &          &         &          \\
DIB      & 5494.10&    0.93  &$-$18     &         &         &         &          &         &          \\
FeII(55) & 5534.84&0.97/1.14 &$-$215:/$-$39&      &         &         &          &         &          \\
DIB      & 5762.70&    0.97  &$-$20:    &         &         &         &          &         &          \\
DIB      & 5766.16&    0.97  &$-$18:    &         &         &         &          &         &          \\
DIB      & 5780.37&    0.77  &$-$10     &0.79     &$-$16    &0.78     &$-$9      &0.78     &$-$8      \\
DIB      & 5796.96&    0.80  &$-$10     &0.79     &$-$15:   &0.79     &$-$10:    &0.81     &$-$10     \\
DIB      & 5809.24&    0.96  &$-$17:    &         &         &         &          &         &          \\
DIB      & 5849.80&    0.92  &$-$17:    &0.94     &$-$17    &         &          &0.94     &$-$18     \\
HeI(11)  & 5875.72&    1.08  &$-$111    &1.12     &$-$92    &1.10     &$-$131    &1.08     &$-$119    \\
NaI(1)   & 5889.95&    0.47  &$-$195    &         &         &0.72     &$-$247    &0.76/0.83&$-$473/$-$309\\
         &        &0.64/1.22 &$-$155/$-$17&0.65/0.92&$-$221/$-$137&0.63/0.56&$-$156/$-$91  &0.73/0.68&$-$222/$-$126\\
         &        &    0.02  &$-$18     &0.10/0.11&-20/-4   &0.03/0.01&$-$23/$-$6&0.00/0.01&$-$19/$-$4\\
NaI(1)   & 5895.92&    0.63  &$-$196    &         &         &0.84     &$-$237    &         &          \\
         &        &0.80/1.17 &$-$150:/$-$14&0.83/0.97&$-$224/$-$130&0.77/0.63&$-$162/$-$93 &0.68 &$-$126\\
         &        &    0.03  &$-$18     &0.08/0.10&$-$19/$-$5&0.03/0.01&$-$20/$-$5&$-$/0.03&$-$/$-$17   \\
FeII(46) & 5991.37&$-$/1.08  &$-$/$-$46:&         &         &         &          &         &          \\
FeII(74) & 6147.74&0.96/1.10 &          &         &         &         &          &         &          \\
FeII(74) & 6149.25&          &          &         &         &         &          &         &          \\
DIB      & 6195.96&     0.86 &$-$13     &         &         &0.92     &$-$15     &0.88     &$-$15     \\
DIB      & 6203.08&     0.89 &$-$12     &         &         &0.89     &$-$16     &0.91     &$-$25     \\
FeII(74) & 6238.39&0.97/1.10 &$-$218:/$-$38:&     &         &$-$/1.07 &$-$/$-$35 &         &          \\
FeII(74) & 6239.95&          &          &         &         &         &          &         &          \\
FeII(74) & 6247.55&0.97/1.21 &$-$203:/$-$27&      &         &0.97/1.15&$-$220:/$-$29&$-$/1.18&$-$/$-$26\\
DIB      & 6269.75&     0.90 &$-$13     &         &         &0.90     &$-$13     &0.90     &$-$13     \\
DIB      & 6283.85&     0.76 &$-$9:     &         &         &         &          &         &          \\
$[OI]1F$ & 6300.30&     1.24 &$-$38     &         &         &1.31     &$-$35     &1.31     &$-$35     \\
$[OI]1F$ & 6363.78&     1.07 &$-$40     &1.07     &$-$44    &1.10     &$-$41     &1.09     &$-$43     \\
         &        &          &          &1.32     &0        &         &          &         &          \\
DIB      & 6375.95&     0.91 &$-$16:    &         &         &         &          &0.94     &$-$10     \\
DIB      & 6379.29&     0.81 &$-$15     &0.81     &$-$16    &         &          &0.84     &$-$13     \\
FeII(74) & 6416.93&0.97/1.08 &$-$/$-$29:&         &         &         &          &         &          \\
FeII(40) & 6432.70&0.97/1.09 &$-$167:/$-$38&      &         &         &          &$-$/1.11&$-$/$-$28  \\
FeII(74) & 6456.38&0.96/1.24 &$-$212/$-$29&       &         &         &          &0.97/1.24&$-$321/$-$17\\
FeII(40) & 6516.08&          &          &         &         &         &          &         &          \\
H$\alpha$& 6562.81&     1.07 &$-$376    &         &         &         &          &         &          \\
         &        &     0.47 &$-$256    &0.44     &$-$270   &0.46     &$-$275    &0.84/0.33&$-$657/$-$326\\
         &        &0.80:/14.2&$-$155:/27&10.8     &26       &13.4     &35        &12.9     &$-$2      \\
DIB      & 6613.56&     0.78 &$-$15     &0.79     &$-$17    &0.78     &$-$16     &0.79     &$-$15     \\
DIB      & 6660.64&     0.91 &$-$12     &         &         &         &          &         &          \\
LiI(1)   & 6707.80&          &          &0.97     &$-$4     &0.97     &$-$14     &         &          \\
HeI(46)  & 6678.15&          &          &         &         &1.05     &$-$137    &1.03     &$-$106    \\
FeII(73) & 7462.38&          &          &         &         &1.04     &$-$33     &1.05     &$-$33     \\
Pa 29    & 8292.31&          &          &         &         &1.02     &$-$59     &         &          \\
Pa 20    & 8392.40&          &          &1.21     &$-$39    &1.24     &$-$39     &1.22     &$-$48     \\
Pa 19    & 8413.31&          &          &1.24     &$-$29    &1.23     &$-$37     &1.22     &$-$41     \\
CaII(2)  & 8542.09&          &          &3.21     &$-$32    &2.72     &$-$17     &3.73     &$-$26     \\
Pa 14    & 8598.39&          &          &1.43     &$-$45    &1.47     &$-$35     &1.43     &$-$42     \\
\noalign{\smallskip}\tableline
\end{tabular}
} \addtocounter{table}{-1}
\end{table}

\begin{table}[H]
\caption{Line identification, intensities and radial velocities
$V_r$ in the optical spectrum of IRAS\,00470+6429. Continued} {\tiny
\begin{tabular}{lccccccccc}
\noalign{\smallskip}\tableline\noalign{\smallskip}
Line     & $\lambda_{\rm lab}$& \multicolumn{2}{c}{11/14/2005} & \multicolumn{2}{c}{12/18/2005} & \multicolumn{2}{c}{12/27/2006}&\multicolumn{2}{c}{12/14/2008}\\
\cline{3-4}\cline{5-6}\cline{7-8}\cline{9-10}
         &  \AA   &   $r$    &  $V_r$   &   $r$   &  $V_r$  &  $r$    & $V_r$    &  $r$    & $V_r$    \\
\noalign{\smallskip}\tableline\noalign{\smallskip}
DIB      & 8620.50&          &          &0.96     &$-$10:   &0.96     &$-$10:    &         &          \\
NI(7)    & 8629.24&          &          &         &         &1.11     &$-$48     &         &          \\
Pa 12    & 8750.47&          &          &         &         &1.57     &$-$35     &1.54     &$-$39     \\
Pa 10    & 9014.91&          &          &         &         &1.50     &$-$30     &         &          \\
Pa 9     & 9229.00&          &          &         &         &1.52     &$-$35     &         &          \\
Pa  7    &10049.40&          &          &1.88     &$-$48    &1.93     &$-$20     &2.00     &$-$31     \\
\noalign{\smallskip}\tableline
\end{tabular}
} \tablecomments{Column 1: line ID, multiplet number is given in
brackets; Column 2: line rest wavelength, the remaining column pairs
contain residual intensities of the absorption/emission peak in
continuum units ($r$) and corresponding heliocentric radial
velocities $V_r$ in km\,s$^{-1}$ for the date specified in the
column head. A hyphen means that the line component is either not
recognized due to noise or its parameters can hardly be measured. If
parameters of a line were hard to measure in a particular spectrum,
no information about this line was entered in the Table.}
\end{table}

\begin{table}[H]
\caption{Heliocentric radial velocities of representative lines in
the spectra of stars close to IRAS\,00470+6429 on the sky}\label{t3}
{\small
\begin{tabular}{llcllcccc}
\noalign{\smallskip}\tableline\noalign{\smallskip}
Object & Sp.T. & $\rho$ &  Date  & Tel. & D  &\multicolumn{3}{c}{$V_r$\,(km\,s$^{-1}$)}\\
\cline{7-9}
 & & & & & & H {\sc I} & Na {\sc I}(IS)& DIB \\
\noalign{\smallskip}\tableline\noalign{\smallskip}
Hiltner\,62        & B3 {\sc III}          & 0.5 & 03/07/1999 & SAO & 2.3 & $-$4  & $-$62/$-$14         & $-$13\\
HD\,4841           & B5 {\sc Ia}           & 1.0 & 08/11/2006 & SAO & 2.4 & $-$35 & $-$58/$-$36/$-$12   & $-$11\\
HD\,4694           & B3 {\sc Ib}$^{\star}$ & 0.1 & 12/04/2006 & SAO & 2.2 &       & $-$64/$-$14         & $-$13\\
HD\,4694           & B3 {\sc Ib}           & 0.1 & 12/27/2006 & McD & 2.2 & $-$50 & $-$65/$-$13         & $-$14\\
BD\,+63$\arcdeg$87 & B0.5 {\sc IV}         & 0.7 & 10/04/2008 & SPM & 2.1 & $-$86 & $-$18               & $-$18\\
BD\,+63$\arcdeg$102& B1 {\sc III}$^{\star}$& 0.2 & 10/09/2008 & SPM & 2.4 & $-$33 & $-$63/$-$13         & $-$16\\
Hiltner\,74        & B2 {\sc V}            & 0.3 & 12/13/2008 & McD & 2.3 &       & $-$16               & $-$15\\
\noalign{\smallskip}\tableline
\end{tabular}
} \tablecomments{Column 1: object ID; Column 2: MK type from SIMBAD
(stars classified by us are marked with an asterisk); Column 3:
angular distance $\rho$ from IRAS\,00470+6429 in degrees; Column 4:
observing date (MM/DD/YYYY); Column 5: observatory where the
spectrum was obtained (SAO -- Special Astrophysical Observatory, McD
-- McDonald Observatory, SPM -- San Pedro Martir); Column 6:
spectroscopic distance in kpc; Columns 7--9: heliocentric radial
velocities of the photospheric Balmer lines and interstellar
features listed in the table header.}
\end{table}

\begin{table}[H]
\caption{List of lines identified in the near-IR spectrum of
IRAS\,00470+6429}\label{t4}
\begin{tabular}{rllrll}
\noalign{\smallskip}\tableline\noalign{\smallskip}
$\lambda_{\rm lab}$& Line& Rem. & $\lambda_{\rm lab}$ & Line& Rem.  \\
\AA             & ID  &      & \AA              & ID  &       \\
\noalign{\smallskip}\tableline\noalign{\smallskip}
9997.00 &   Fe {\sc II}&         &15191.8 &   Br 20      &         \\
10049.4 &   Pa 7       &         &15260.5 &   Br 19      &         \\
10121.7 &   Ca {\sc I} & weak    &15341.8 &   Br 18      &         \\
10455.4 &   Fe {\sc II}&         &15387.7 &   Ca {\sc I} &         \\
10499.3 &   Fe {\sc II}&         &15438.9 &   Br 17      &         \\
10686.5 &   Fe {\sc II}&         &15556.5 &   Br 16      &         \\
10830.3 &   He {\sc I} & emis/abs&16109.3 &   Br 13      &         \\
10938.1 &   Pa  6      &         &16407.2 &   Br 12      &         \\
11286.8 &   O  {\sc I} &         &16806.5 &   Br 11      &         \\
11748.8 &   Ca {\sc I} &         &16873.2 &   Fe {\sc II}& weak    \\
12818.1 &   Pa  5      &         &21655.3 &   Br-$\gamma$&         \\
\noalign{\smallskip}\tableline
\end{tabular}
\tablecomments{Column 1: line rest wavelength; Column 2: line ID;
Column 3: remarks on the line properties. All the lines listed in
this table are in emission, except the He {\sc I} line.}
\end{table}

\begin{table}[H]
\caption{Observing log of the photometric observations of
IRAS\,00470+6429}\label{t5}
\begin{tabular}{ccccccccc}
\tableline\noalign{\smallskip}
Observatory & Telescope & Filters & JD     & N & Err.\\
            &           &         &2450000+&   & mag \\
\noalign{\smallskip}\tableline\noalign{\smallskip}

Los Alamos  & 0.20--m    & Clear   &1335--1571&180&0.02\\
TSAO        & 0.48--m    & $UBVR$  &2970--2974& 2 &0.03\\
Maidanak    & 0.48--m    & $UBVR$  &3230--3654& 24&0.02\\
LHO         & 0.30--m    & $BVRI$  &3353--4779&140&0.02\\
SRO         & 0.36--m    & $BVRI$  &4778--4831& 18&0.02\\
Persbuhaugen& 0.35--m    & $V$     &4098      & 47&0.01\\
Bossmo      & 0.25--m    & $V$     &4124      &390&0.01\\
Bossmo      & 0.25--m    & Clear   &4460      &282&0.01\\
\noalign{\smallskip}\tableline
\end{tabular}
\tablecomments{Column 1: Observatory name: Los Alamos (New Mexico,
NSVS survey, see text), Tien-Shan (TSAO, Kazakhstan), Maidanak
(Uzbekistan) Lizard Hollow (LHO, Arizona), Sonoita Research
Observatory (SRO, Arizona), Persbuhaugen (Vegglifjell, Norway),
Bossmo (Mo i Rana, Norway); Column 2: telescope size; Column 3:
filters used (no filter was used for the observations at Bossmo on
JD2454460); Column 4: time period of the observations in Julian
dates; Column 5: number of observations obtained; column 6 - typical
uncertainty of individual observations in magnitudes.}
\end{table}

\clearpage

\end{document}